\documentclass[showpacs,aps]{revtex4}
\usepackage{epsfig,amsmath}

\begin{document}
\title{\bf Electrovacuum Static Counterrotating  Relativistic Dust Disks}

\author{Gonzalo Garc\'\i a R.}\email[e-mail: ]{ggr1970@yahoo.com}
\author{Guillermo A. Gonz\'{a}lez}\email[e-mail: ]{guillego@uis.edu.co}

\affiliation{Escuela de F\'{\i}sica, Universidad Industrial de Santander,
A. A. 678, Bucaramanga, Colombia}

\pacs{04.20.-q, 04.20.Jb, 04.40.Nr}

\begin{abstract}

A detailed study is presented of the counterrotating model (CRM) for generic
electrovacuum static axially symmetric relativistic thin disks without radial
pressure. We find a general constraint over the counterrotating tangential
velocities needed to cast the surface energy-momentum tensor of the disk as the
superposition of two counterrotating charged dust fluids. We also find explicit
expressions for the energy densities, charge densities and velocities of the
counterrotating fluids. We then show that this constraint can be satisfied if
we take the two counterrotating streams as circulating along electro-geodesics.
However, we show that, in general, it is not possible to take the two
counterrotating fluids as circulating along electro-geodesics nor take the two
counterrotating tangential velocities as equal and opposite. Four simple
families of models of counterrotating charged disks based on Chazy-Curzon-like,
Zipoy-Voorhees-like, Bonnor-Sackfield-like and Kerr-like electrovacuum
solutions are considered where we obtain some disks with a CRM well behaved.
The models are constructed using the well-known ``displace, cut and reflect''
method extended to solutions of vacuum Einstein-Maxwell equations. 

\end{abstract}

\maketitle

\section{Introduction}

Stationary or static axially symmetric exact solutions of Einstein  equations
describing relativistic thin disks are of great astrophysical importance since
can be used as models of certain stars, galaxies and accretion disks. These
were first studied by Bonnor and Sackfield \cite{BS}, obtaining pressureless
static disks, and then by Morgan and Morgan, obtaining static disks with and
without radial pressure \cite{MM1,MM2}. In connection with gravitational
collapse, disks were first studied by Chamorro, Gregory and Stewart
\cite{CHGS}. Disks with radial tension have been also studied \cite{GL1}.
Several classes of exact solutions of the Einstein field equations
corresponding to static and stationary thin disks have been obtained by
different authors \cite{LP,LO,LEM,LL1,BLK,BLP,BL,LL2,LL3,KLE,GL2}, with or
without radial pressure.

In the case of static disks without radial pressure, there are two common
interpretations. The stability of these models can be explained by either
assuming the existence of hoop stresses or that the particles on the disk plane
move under the action of their own gravitational field in such a way that as
many particles move clockwise as counterclockwise. This last interpretation,
the ``counterrotating model'' (CRM), is frequently made since it can be invoked
to mimic true rotational effects. Even though this interpretation can be seen
as a device, there are observational evidence of disks made of streams of
rotating and counterrotating matter \cite{RGK,RFF}.

Disk sources for stationary axially symmetric spacetimes with magnetic fields
are also of astrophysical importance mainly in the study of neutron stars,
white dwarfs and galaxy formation. Although disks with electric fields do not
have clear astrophysical importance, their study may be of interes in the
context of exact solutions. Thin disk  have been discussed as sources for
Kerr-Newman fields \cite{LBZ}, magnetostatic axisymmetric fields \cite{LET1}
and conformastationary metrics \cite{KBL}. Following the Ref. \cite{LBZ} the
resultating disks can  be interpreted either as rings with internal pressure
and currents or as two counterrotating  streams of freely moving charged
particles, i.e. which move along electro-geodesics (solution to the geodesic
equation in the presence of a Lorentz force).

In all the above cases, the disks are obtained by an ``inverse problem''
approach, called by Synge the ``{\it g-method}'' \cite{SYN}. The method works
as follows: a solution of the vacuum Einstein equations is taken, such that
there is a discontinuity in the derivatives of the metric tensor on the plane
of the disk, and the energy-momentum tensor is obtained from the Einstein 
equations. The physical properties of the matter distribution are then studied
by an analysis of the surface energy-momentum tensor so obtained. On the other
hand, a ``direct problem'' approach, called by Synge  the ``{\it T-method}'',
is also used by other authors \cite{NM,KLE1,KR,KLE2,FK,KLE3,KLE4} by taking a
given surface energy-momentum tensor and solving the Einstein equations in the
matter region. The inner solution is then used to obtain boundary data for the
vacuum field equations in the outer region. The T-method is used in Ref.
\cite{NM} to obtain the solution to the problem of a (one-component) uniformly
rotating disk of dust, and in Refs. \cite{KLE1,KR,KLE2,FK,KLE3,KLE4} to
generate counterrotating  dust disks, but no condition is imposed there about
the (electro-) geodesic motion  of the two counterrotating streams.

The aim of this paper is to perform a detailed study of the CRM for generic 
electrovacuum static axially symmetric relativistic thin disks without radial
pressure. The counterrotating model for the case of static thin disks without
electric or magnetic fields was recently studied in \cite{GE}, so the material
presented here is a continuation of the mentioned work. The paper is organized
as follows. In Sec. II we present a summary of the procedure to obtain thin
disks models  with a purely azimuthal pressure  and currents, using the
well-known ``displace, cut and reflect'' method extended to solutions of 
Einstein-Maxwell equations. In particular, we obtain expressions for the
surface energy-momentum tensor and the current density of the disks.  

Next, in Sec. III, the disks are interpreted in terms of the CRM. We find a
general constraint over the counterrotating tangential velocities needed to
cast the surface energy-momentum tensor of the disk as the superposition of two
counterrotating  charged dust fluids. We also find explicit expressions for the
energy densities, current densities and velocities of the counterrotating
fluids. We then show that this constraint can be satisfied if we take the two
counterrotating streams as circulating along electro-geodesics. However, we show
that, in general, it is not possible to take the two counterrotating fluids as
circulating along electro-geodesics nor take the two counterrotating tangential
velocities as equal and opposite.

In the following section, Sec. IV, four simple families of models of
counterrotating charged disks based on Chazy-Curzon-like, Zipoy-Voorhees-like,
Bonnor-Sackfield-like and Kerr-like metrics are presented where we obtain some
disks with a CRM well behaved. In particular, we study the tangential
velocities, mass and electric charge densities of both streams in the pure
static and electrostatic (magnetostatic) cases. Also the stability against
radial perturbation is analyzed. Finally, in Sec. V, we summarize our main
results.   

\section{Electrovacuum Static Relativistic Disks}

In this section we present a summary of the procedure to obtain electrovacuum 
static axially symmetric relativistic thin disks. The  simplest metric  to
describe a  static axially symmetric spacetime is the Weyl's line element 
\begin{equation}
ds^2 = - \ e^{2 \nu} dt^2 \ + \ e^{- 2 \nu} [r^2 d\varphi^2 + e^{2 \lambda}
(dr^2 + dz^2)] , \label{eq:met}
\end{equation}
where $\nu$ and $\lambda$  are functions of $r$ and $z$ only.  The vacuum
Einstein-Maxwell equations, in  geometrized units such that $8 \pi G = c = \mu
_0 = \varepsilon _0 = 1$,  are given by 
\begin{subequations}
\begin{eqnarray}
&   &    R_{ab} \  =  \ T_{ab},  \\
&   &     \nonumber       \\
&   &     T_{ab} \  =  \ F_{ac}F_b^{ \ c} - \frac 14 g_{ab}F_{cd}F^{cd},  \\
&   &     \nonumber    \\
&   &    F^{ab}_{ \ \ \ ; b} = 0,    \\
&   &     \nonumber       \\
&   &   F_{ab} =  A_{b,a} -  A_{a,b},
\end{eqnarray}\label{eq:einmax}\end{subequations} 
where all symbols are understood.

For the metric (\ref{eq:met}), the  Einstein-Maxwell equations in vacuum are
equivalent  the  complex Ernst equations  \cite{E2}
\begin{subequations}
\begin{eqnarray}
f \Delta {\cal E} &=& (\nabla {\cal E} + 2\Phi^\ast\nabla\Phi) \cdot
\nabla{\cal E}, \label{eq:ece1}     \\
&  & \nonumber  \\
f \Delta \Phi &=& (\nabla {\cal E} + 2\Phi^\ast\nabla\Phi) \cdot \nabla\Phi,
\label{eq:ece2} 
\end{eqnarray}\label{eq:ece}\end{subequations}
with ${\cal E}={\cal E}^*$ (static spacetime), where  $\Delta$ and  $\nabla$
are  the  standard differential operators in cylindrical coordinates and $f=
e^{2\nu}$. The metric functions are obtained via 
\begin{subequations}\begin{eqnarray}
f  &=& {\cal E} +\Phi \Phi^*  ,  \\
&&	\nonumber	\\
\lambda _{,\zeta} &=&  \frac{\sqrt{2} r}{4 f^2} ( {\cal E}_{,\zeta} +2 \Phi ^*
\Phi_{,\zeta} )({\cal E}_{,\zeta} +2 \Phi  \Phi^*_{,\zeta} ) 
 - \ \frac{\sqrt{2} r}{f}\Phi_{,\zeta}\Phi^*_{,\zeta} ,
\label{eq:Lam} 
\end{eqnarray}\end{subequations}
where $\sqrt 2\zeta = r +iz$, so that $\sqrt 2 \partial_{,\zeta}=
\partial_{,r}-i\partial_{,z}$. The electromagnetic potencials are related to 
$\Phi$ via
\begin{subequations}\begin{eqnarray}
A_t &=& \sqrt 2{\rm Re} \Phi , \\
&& \nonumber \\
A_{\varphi,\zeta} &=&  \sqrt 2 i \frac r f ( {\rm Im} \Phi )_{,\zeta} .
\end{eqnarray}\end{subequations}

In order to obtain a solution of (\ref{eq:ece})  representing a thin disk at
$z=0$, we assume that the components of the metric tensor are continuous across
the disk, but their first derivates discontinuous on the plane $z=0$, with
discontinuity functions
$$
b_{ab} \ = g_{ab,z}|_{_{z = 0^+}} \ - \ g_{ab,z}|_{_{z = 0^-}} \ =
\ 2 \ g_{ab,z}|_{_{z = 0^+}} .
$$
Thus, the Einstein-Maxwell equations yield an energy-momentum tensor $T_a^b \ =
\ Q_a^b \ \delta (z)$ and a planar current density ${\rm J}_a = j_a  \delta (z)
= - 2 A_{a,z} \delta (z)$, where $\delta (z)$ is the usual Dirac function with
support on the disk and
$$
Q^a_b = \frac{1}{2}\{b^{az}\delta^z_b - b^{zz}\delta^a_b + g^{az}b^z_b -
g^{zz}b^a_b + b^c_c (g^{zz}\delta^a_b - g^{az}\delta^z_b)\}
$$
is the distributional energy-momentum tensor. The ``true'' surface
energy-momentum tensor (SEMT) of the disk, $S_a^b$, can be obtained through the
relation
\begin{equation}
S_a^b \ = \ \int T_a^b \ ds_n \ = \ e^{\lambda - \nu} \ Q_a^b ,
\end{equation}
where $ds_n = \sqrt{g_{zz}} \ dz$ is the ``physical measure'' of length in the
direction normal to the disk, and the surface current density as ${\rm j}_a= \
e^{\lambda -\nu} j_a $. For the metric (\ref{eq:met}), the  non-zero 
components of  $S_a^b$ and the current density are
\begin{subequations}\begin{eqnarray}
&S^0_0 &= \ 2 e^{\nu - \lambda} \left\{ \lambda,_z - \ 2 \nu,_z  \ \right\} ,
 \label{eq:emt1}     		\\
&	&	\nonumber	\\
&S^1_1 &= \ 2 e^{\nu - \lambda} \lambda,_z , \label{eq:emt2}
\end{eqnarray}\label{eq:emt}\end{subequations}
and
\begin{subequations}\begin{eqnarray}
& {\rm j}_t &= \ -2 e^{\nu - \lambda} A _{t,z} , \label{eq:corelec}   \\
&	&	\nonumber	\\
&{\rm j}_{\varphi} &= \ -2 e^{\nu - \lambda} A _{\varphi,z},  \label{eq:cormag}
\end{eqnarray}\label{eq:cor}\end{subequations}
where all the quantities are evaluated at $z = 0^+$.

With an orthonormal tetrad ${{\rm e}_{\hat a}}^b = \{ V^b , W^b , X^b , Y^b
\}$, where
\begin{subequations}\begin{eqnarray}
V^a &=& e^{- \nu} \ ( 1, 0, 0, 0 ) ,	\\
	&	&	\nonumber	\\
W^a &=& \frac{e^\nu} {r} \ \ ( 0, 1, 0, 0 ) ,	\\
	&	&	\nonumber	\\
X^a &=& e^{\nu - \lambda} ( 0, 0, 1, 0 ) ,	\\
	&	&	\nonumber	\\
Y^a &=& e^{\nu - \lambda} ( 0, 0, 0, 1 ) ,
\end{eqnarray}\label{eq:tetrad}\end{subequations}
we can write the metric and the SEMT in the canonical forms
\begin{subequations}\begin{eqnarray}
&	&g_{ab} \ = \ - V_a V_b + W_a W_b + X_a X_b + Y_a Y_b ,
\label{eq:metdia}							\\
&	&	\nonumber						\\
&	&S_{ab} \ = \ \epsilon V_a V_b + p_\varphi W_a W_b  \
, \label{eq:emtdia}
\end{eqnarray}\end{subequations}
where
\begin{equation}
\epsilon \ = \ - S^0_0 \quad , \quad p_\varphi \ = \ S^1_1 \quad  ,
\label{eq:dps}
\end{equation}
are, respectively, the energy density and the azimuthal pressure of the disk. 

\section{The counterrotating model}

We now consider, based on references \cite{LET2} and \cite{FMP}, the
possibility that the SEMT $S^{ab}$ and the current density  ${\rm j}^a$ can be
written as the superposition of two counterrotating  fluids that circulate in
opposite directions; that is, we assume 
\begin{subequations}\begin{eqnarray}
S^{ab} &=& S_+^{ab} \ + \ S_-^{ab} , \label{eq:emtsum}   \\
       & &	        \nonumber	                    \\
{ \rm j}^a    &=& { \rm j}_+^a   +{\rm j}_-^a, \label {eq:corsum} 
\end{eqnarray}\end{subequations}
where the  quantities in the right-hand side  are, respectively, the SEMT and
the current density of the prograde and retrograde counterrotating fluids. 

Let  $U_\pm^a = ( U_\pm^0 , U_\pm^1, 0 , 0 )$ be the velocity vectors of the
two counterrotating fluids. In order to do the decomposition (\ref{eq:emtsum})
and (\ref{eq:corsum}) we project the velocity vectors onto the tetrad ${{\rm
e}_{\hat a}}^b$, using the relations \cite{CHAN}
\begin{equation}
U_\pm^{\hat a} \ = \ {{\rm e}^{\hat a}}_b U_\pm^b \qquad , \qquad U_\pm^ a = \
U_\pm^{\hat c} {{\rm e}_{\hat c}}^a  .
\end{equation}
With the tetrad (\ref{eq:tetrad}) we can write
\begin{equation}
U_\pm^a \ = \ \frac{ V^a + {\rm U}_\pm W^a }{\sqrt{1 - {\rm U}_\pm^2}} ,  
\label{eq:vels}
\end{equation}
and thus
\begin{subequations}\begin{eqnarray}
&V^a &= \ \frac{\sqrt{1 - {\rm U}_-^2} {\rm U}_+ U_-^a - \sqrt{1 - {\rm U}_+^2}
{\rm U}_- U_+^a}{{\rm U}_+ - {\rm U}_-} , \label{eq:va} \\
&	&	\nonumber	\\
&W^a &= \ \frac{\sqrt{1 - {\rm U}_+^2} U_+^a - \sqrt{1 - {\rm U}_-^2}
U_-^a}{{\rm U}_+ - {\rm U}_-} , \label{eq:wa}
\end{eqnarray}\label{eq:vawa}\end{subequations}
where ${\rm U}_\pm = U_\pm^{\hat 1} / U_\pm^{\hat 0}$ are the tangential
velocities of the fluids with respect to the tetrad.

Using (\ref{eq:vawa}), we can write the SEMT as
\begin{eqnarray}
S^{ab} &=& \frac{ f( {\rm U}_- , {\rm U}_- ) (1 - {\rm U}_+^2) \ U_+^a U_+^b
}{({\rm U}_+ - {\rm U}_-)^2} +  \frac{ f( {\rm U}_+ , {\rm U}_+ ) (1 - {\rm
U}_-^2) \ U_-^a U_-^b }{({\rm U}_+ - {\rm U}_-)^2}	\nonumber	\\
&	&		\nonumber	\\
& & - \frac{ f( {\rm U}_+ , {\rm U}_- ) [(1 - {\rm U}_+^2)(1 - {\rm
U}_-^2)]^{\frac{1}{2}} ( U_+^a U_-^b + U_-^a U_+^b ) }{({\rm U}_+ - {\rm
U}_-)^2},	\nonumber	
\end{eqnarray}
where
\begin{equation}
f( {\rm U}_1 , {\rm U}_2 ) \ = \  \epsilon  {\rm U}_1 {\rm U}_2 + p_\varphi  \
. \label{eq:fuu}
\end{equation}

Clearly, in order to cast the SEMT in the form (\ref{eq:emtsum}), the mixed
term must be absent and therefore the counterrotating tangential velocities
must be related by
\begin{equation}
f( {\rm U}_+ , {\rm U}_- ) \ = \ 0 , \label{eq:liga}
\end{equation}
where we assume that $|{\rm U}_\pm| \neq 1$. Then, assuming a given choice for
the counterrotating velocities in agreement with the above relation, we can
write the SEMT as (\ref{eq:emtsum}) with
\begin{equation}
S^{ab}_\pm =  \epsilon_\pm  \ U_\pm^a U_\pm^b ,
\end{equation}
so that we have two counterrotating dust fluids with  energy densities given
by
\begin{equation}
\epsilon_\pm  = \left[ \frac{ 1 - {\rm U}_\pm ^2 }{{\rm U}_\mp - {\rm U}_\pm}
\right] {\rm U}_\mp \epsilon. \label{eq:enercon} 
\end{equation}
Thus the SEMT $S^{ab}$ can be written as the superposition of two
counterrotating dust streams if, and only if, the constraint (\ref{eq:liga}) 
admits a solution such that ${\rm U}_+ \neq {\rm U}_-$. This result is
completely equivalent to the necessary and sufficient condition obtained in
reference \cite{FMP}.

Similarly, we can write the current density as (\ref{eq:corsum}) with
\begin{equation}
{\rm j}^a_\pm  = \sigma _\pm U_\pm^a, 
\end{equation}
where $\sigma _\pm$ are the counterrotating electric charge densities of the
fluids which are given by
\begin{equation}
\sigma _{ \pm} =  \left[\frac { \sqrt{1-{\rm U}^2_\pm }} {{  {\rm U}_\pm -\rm
U}_\mp}\right] ( \frac {\rm j^1}{W^1} -\frac {\rm j^0}{V^0}{\rm U}_\mp  ).
\label{eq:sig} 
\end{equation}

Another quantity related with the counterrotating motion is the specific
angular momentum of a particle rotating at a radius $r$, defined as $h_\pm =
g_{\varphi\varphi} U_\pm^\varphi$. We can write
\begin{equation}
h_\pm \ = \ \frac{r e^{- \nu} {\rm U}_\pm}{\sqrt{1 - {\rm U}_\pm^2}} .
\label{eq:moman}
\end{equation}
This quantity can be used to analyze the stability of the disks against radial
perturbations. The condition of stability,
\begin{equation}
\frac{d(h^2)}{dr} \ > \ 0 ,
\end{equation}
is an extension of the Rayleigh criteria of stability of a fluid in rest in a
gravitational field \cite{FLU}.

As we can see from Eqs. (\ref{eq:vels}), (\ref{eq:enercon}), (\ref{eq:sig}) and
(\ref{eq:moman}), all the physical quantities asociated with the CRM depend of
the counterrotating tangential velocities ${\rm U}_\pm$. However, the
constraint (\ref{eq:liga}) does not determine ${\rm U}_\pm$ uniquely, and so we
need to imposse some additional requeriment in order to obtain a complete
determination of the tangential velocities, leading so to a well defined CRM. A
possibility, commonly assumed, is to take the two counterrotating fluids as
circulating along electro-geodesics
\begin{equation}
\frac 12 \epsilon _\pm g_{ab,r}U^a_\pm U^b_\pm = - \sigma _\pm F_{ra} U^a_\pm.
\label{eq:geo}
\end{equation}
The conservation laws (the Bianchi identities) at the disk then imply that the
two counterrotating streams interact only by means of gravitational and
electromagnetic forces.

Let $\omega_\pm = U_\pm^1/U_\pm^0$ be the angular velocities of the particles.
In terms of $\omega_\pm$ we get
\begin{equation}
{\rm U}_\pm \ =  \  \left[ \frac{V^0}{W^1} \right] \omega_\pm ,
\end{equation}
and so, using  (\ref{eq:vels}), (\ref{eq:enercon}) and (\ref{eq:sig}),
(\ref{eq:geo}) takes the form 
$$
\omega_\mp (g_{11,r}\omega_\pm ^2+g_{00,r})=   -  \frac {2 V_0^2} {\epsilon}
({\rm j}^0 \omega_\mp - {\rm j}^1) (A_{t,r} + A_{\varphi,r}\omega_\pm ),
$$
so that
\begin{subequations}\begin{eqnarray}
\omega _+ + \omega _-  &=&  \frac 2 \epsilon \left[ \frac {W^1}{V^0} 
\right]^2 \left[ \frac {\epsilon {\rm j}_1 A_{t,r}+ p_ \varphi {\rm j}_0
A_{\varphi,r} }{\epsilon g_{00,r} +2{\rm j}_0  A_{t,r}} \right],  \\
& &		\nonumber	\\
\omega _+  \omega _- &=& - \left[ \frac {W^1}{V^0} \right]^2 \frac
{p_{\varphi}}{\epsilon}    ,
\end{eqnarray}\end{subequations} 
where we have used  the Einstein-Maxwell equation (\ref{eq:Lam}) and the
expressions (\ref{eq:emt1}) - (\ref{eq:cormag}) for the SEMT and the  current
density. From the last expression follows immediately that $f( {\rm U}_+ , {\rm
U}_- )$ vanishes and so the electro-geodesic velocities agree with
(\ref{eq:liga}) and we have a well defined CRM.

Note that in general the two fluids circulate with different velocities.
However, when the spacetime is electrostatic (or magnetostatic) the two
electro-geodesic fluids circulate with equal and opposite velocities, so that
the constraint (\ref{eq:liga}) is equivalent to
\begin{equation}
{\rm U}^2 \ = \ \frac{p_\varphi}{\epsilon} ,   \label{eq:vel} 
\end{equation}
as is commonly assumed in the works concerning counterrotating disks. We now
have two counterrotating charged dust streams with equal energy densities 
\begin{equation}
\epsilon_\pm \ = \ \frac{\epsilon - p_\varphi}{2},
\end{equation}
specific angular momenta
\begin{equation}
h_\pm = \pm r e^{-\nu} \sqrt { \frac {p_\varphi}{\epsilon - p_\varphi}},
\label{eq:hpm}
\end{equation}
and electric charge densities  
\begin{subequations}\begin{eqnarray}
& \sigma _{e \pm} & = -\frac12 e^{-\nu}{\rm j}_0 \sqrt{1-\frac
{p_\varphi}{\epsilon}},          \label{eq:sim} \\  
&  &     \nonumber    \\
& \sigma _{m \pm} & = \pm \frac {1}{2r} e^{\nu}{\rm j}_1 \sqrt{ \frac
{\epsilon}{p_\varphi}-1 },    \label{eq:sie}  
\end{eqnarray}\end{subequations}
where  $\sigma _{e \pm}$ and $\sigma _{m \pm}$ are the electric charge 
densities in the electrostatic and magnetostatic cases, respectively. The
velocities are given by (\ref{eq:vel}). 

Although the electro-geodesic choice leads to a well defined CRM, sometimes the
corresponding physical quantities may have unphysical behavior. For instance,
the expressions (\ref{eq:sig}) and (\ref{eq:moman}) may take imaginary values.
So we need to consider other solutions of Eq. (\ref{eq:liga}) differents of the
electro-geodesic velocities. Another possibility, commonly considered, is to
take the two counterrotating fluids not circulating along electro-geodesics but
with equal and opposite tangential velocities,
\begin{equation}
{\rm U}_{\pm} = \pm  {\rm U}.
\end{equation}
This choice, that imply the existence of additional interactions between  the
two streams (e.g. colllisions), leads to a complete determination of the
velocity vectors in such away that the expressions for the  velocities, energy 
densities and specific angular momenta of both streams coincide with  the
electrostatic (magnetostatic) case, but the charge densities are differents.
However, this can be made only when $0 \leq |p_\varphi/\sigma| \leq 1$. If this
is not the case, we cannot take the two velocities as equal and opposite. In
the general case, the two counterrotating streams circulate with different
velocities and we can write (\ref{eq:liga}) as
\begin{equation}
{\rm U}_+ {\rm U}_- = - \frac {p_\varphi}{\epsilon}.
\end{equation}
However, this relation does not determine completely the tangential velocities,
and so the CRM is undetermined.

\section{Some Simple Examples of counterrotating Charged Dust Disks}

\subsection{CRM for Chazy-Curzon-like disks}

The first  family of  solutions  considered is a Chazy-Curzon-like solution
which is given by
\begin{subequations}\begin{eqnarray}
e^\nu    &=& \frac {2}{(1+a) e^{\gamma/\rho} +  (1-a) e^{-\gamma/\rho }} , \\
        & & \nonumber   \\
\lambda &=&  - \frac{\gamma^2 \sin^2 \theta}{2\rho^2} , \\
        & & \nonumber   \\
A_t    &=&  \frac{\sqrt{2} p [e^{\gamma/\rho} - e^{-\gamma/\rho}]}{(1+a)
e^{\gamma/\rho} + (1-a) e^{-\gamma/\rho}} , \\
        & & \nonumber   \\
A_\varphi & = &\sqrt 2 \gamma q \cos \theta,
\end{eqnarray}\label{eq:ccl}\end{subequations}
where $a^2=1+b^2$, with  $b^2 = p^2 + q^2 $, and $\gamma$ is a real constant.
Here $p$ and $q$ are the  electric and magnetic parameters, respectively.
$\rho$ and $\theta$ are the spherical coordinates, related to the Weyl
coordinates by 
\begin{equation}
r   = \rho \sin \theta ,  \quad  \quad z +z_0  = \rho \cos \theta.
\end{equation}
Note that we have displaced the origin of the $z$ axis in $z_0$.   This
solution can be generated, in these coordinates,  using the well-known  complex
potencial formalism proposed by Ernst \cite{E2} from the   Chazy-Curzon vacuum
solution \cite{CH,C}, by choosing the parameter $q$ of Ref. \cite{ E2} as
complex. For  $q=0$ we have an electrostatic solution and for $p=0$  one 
obtains its  magnetostatic analogue \cite{B1}.

From the above expressions we can compute the physical quantities associated
with the disks. We obtain 

\begin{subequations}\begin{eqnarray}
\epsilon &=&  \frac { 8\gamma \cos \theta e^{-\lambda} [(\rho -\gamma
\sin^2\theta)(1+a) e^{\gamma /\rho} - (\rho +\gamma \sin^2 \theta) (1-a)
e^{-\gamma/\rho}] } {\rho^3 [(1+a) e^{\gamma/\rho} +  (1-a) e^{-\gamma/\rho}]^2
}  , \\
& &	\nonumber	\\
p_\varphi &=&  \frac { 8\gamma^2 \sin ^2 \theta \cos \theta e^{-\lambda}  }
{\rho^2[(1+a) e^{\gamma/\rho} +  (1-a) e^{-\gamma/\rho}] }  , \\
& &	\nonumber	\\
{\rm j}_t &=&  \frac { 16 \sqrt2  \gamma p  \cos \theta e^{-\lambda}  } {(1+a)
\rho^2 [(1+a) e^{\gamma/\rho} +  (1-a) e^{-\gamma/\rho}]^3 }  , \\
& &	\nonumber	\\
{\rm j}_\varphi &=& - \frac { 8 \sqrt2  \gamma q  \sin ^2 \theta e^{-\lambda} 
} {\rho[(1+a) e^{\gamma/\rho} +  (1-a) e^{-\gamma/\rho}] }  . 
\end{eqnarray}\end{subequations}

\begin{figure*}
$$
\begin{array}{cc}
\epsilon, \  p_\varphi  &  {\rm j}_t, \ {\rm j}_\varphi \\
\epsfig{width=2.5in,file=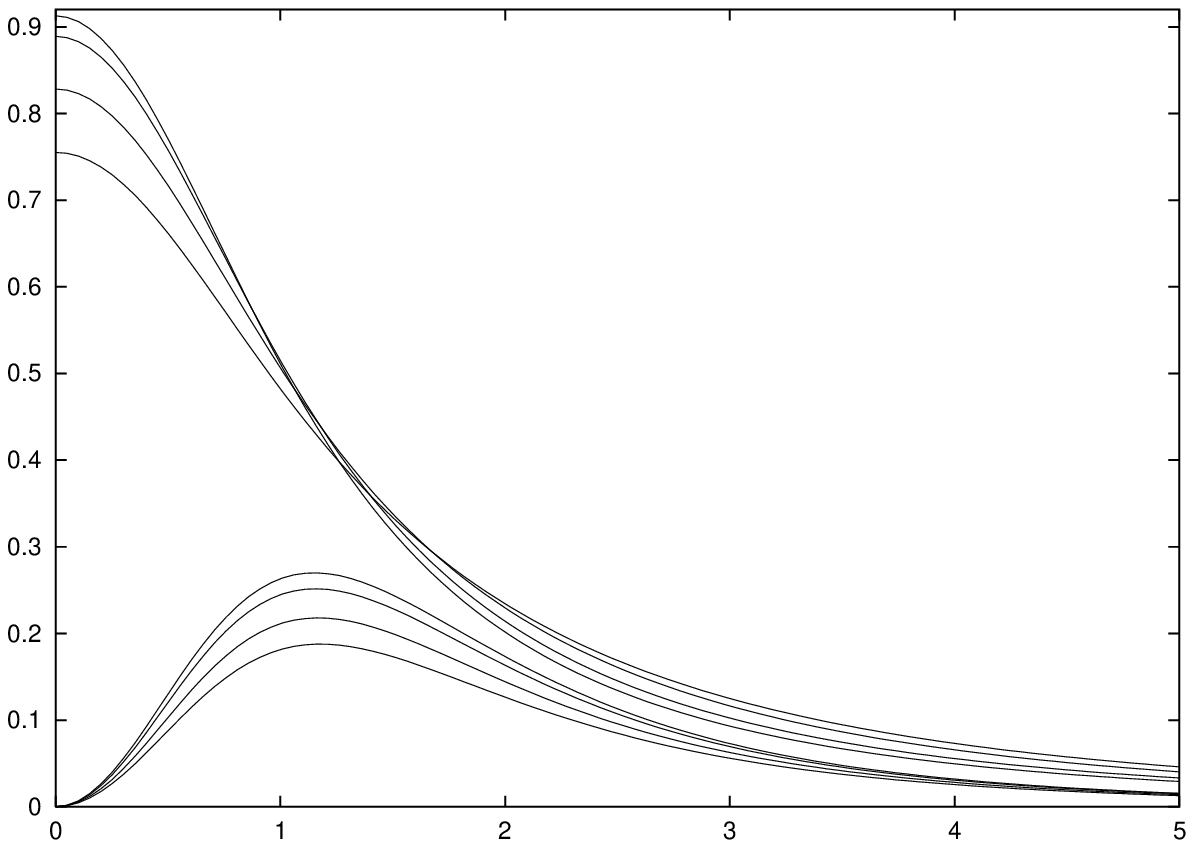} & \epsfig{width=2.5in,file=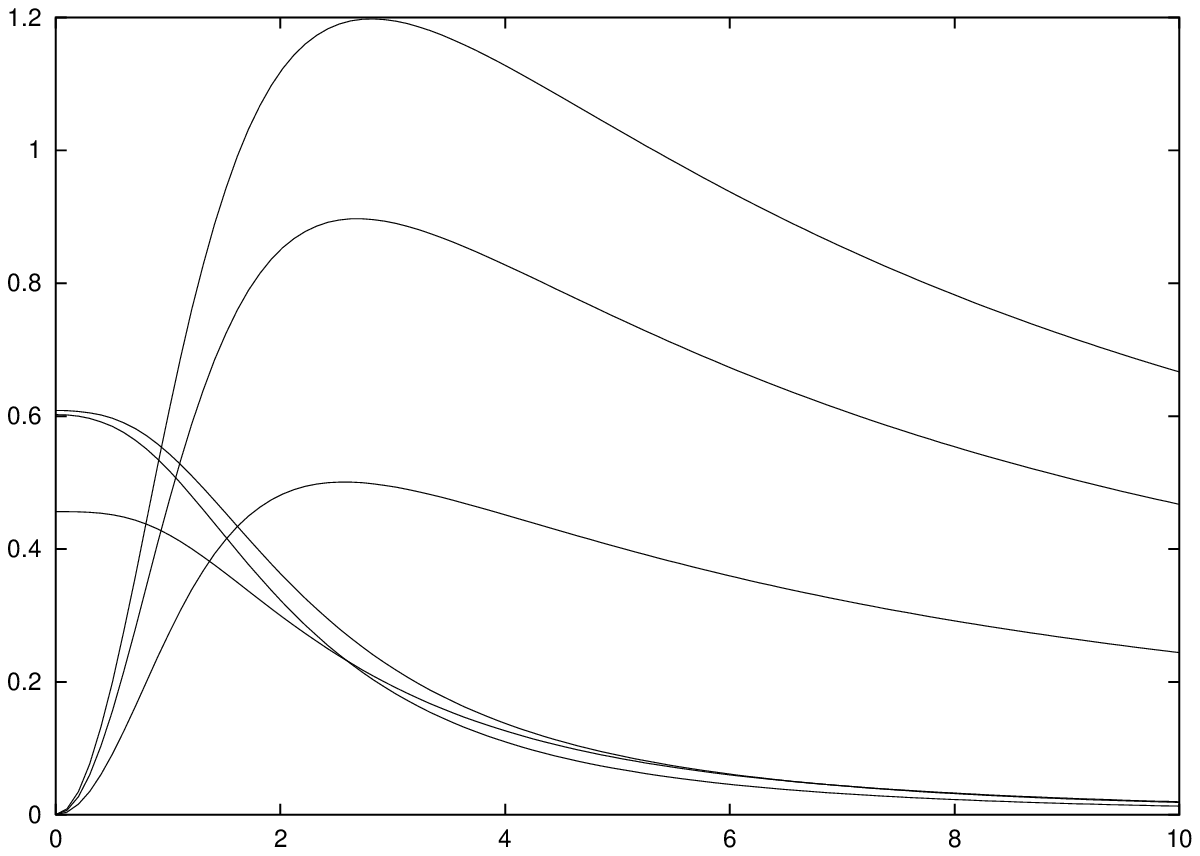} \\
 r      &  r    \\
(a)     &   (b)
\end{array}
$$	
\caption{$(a)$ $\epsilon$ (upper curves) and   $ p_\varphi$ (scaled by a factor
of 2.5),  as functions of $r$ for  Chazy-Curzon-like  disks  with $\gamma=1$,
$z_0=1.5$ and $p=q=0$ (top curves), $0.5$, $1.0$, and $1.5$ (bottom curves). 
$(b)$ ${\rm j}_t$ (lower curves  scaled by a factor of 20) and ${\rm
j}_\varphi$ for the same values of the parameters. }\label{fig:cepco}
\end{figure*}

In order to study the behavior of these quantities we perform a graphical
analysis of them for disks  with $\gamma=1$, $z_0=1.5$ and $p=q=0$, $0.5$,
$0.1$, and $1.5$.  In Fig. \ref{fig:cepco}$(a)$ we show  the energy density 
$\epsilon$ (upper curves) and the azimuthal pressure $p_\varphi$ (lower
curves), as  functions of $r$. We  see  that the energy density presents a
maximum at $r = 0$ and then decreases rapidly with  $r$. We also see that  the
presence of electromagnetic  field decreases the energy density  at the central
region of the disk and  later increases it.  We can observe that the pressure
increases rapidly as one moves away from the disk center, reaches a maximum
and  later  decreases rapidly. We also observe that the electromagnetic field
decreases the pressure everywhere on the disk. Next, the charge and electric
current densities $ {\rm j}_t$  and ${\rm j}_\varphi$ are  represented in Fig.
\ref{fig:cepco}$(b)$.  $ {\rm j}_t$ has a maximum at the disk center and then
falls   to zero at infinity, whereas $ {\rm j}_\varphi$   exhibits a similar
behavior to the pressure. We also cumputed the functions $\epsilon$,
$p_\varphi$, $ {\rm j}_t$ and $ {\rm j}_\varphi$ for other values of the
parameters and, in all the cases, we found  the same behavior.

\begin{figure*}
$$
\begin{array}{cc}
{\rm U}_+, \ \ -{\rm U}_-, \ \   \rm U^2 &   h^2, \ \  h^2_+, \ \  h^2_-   \\
\epsfig{width=2.5in,file=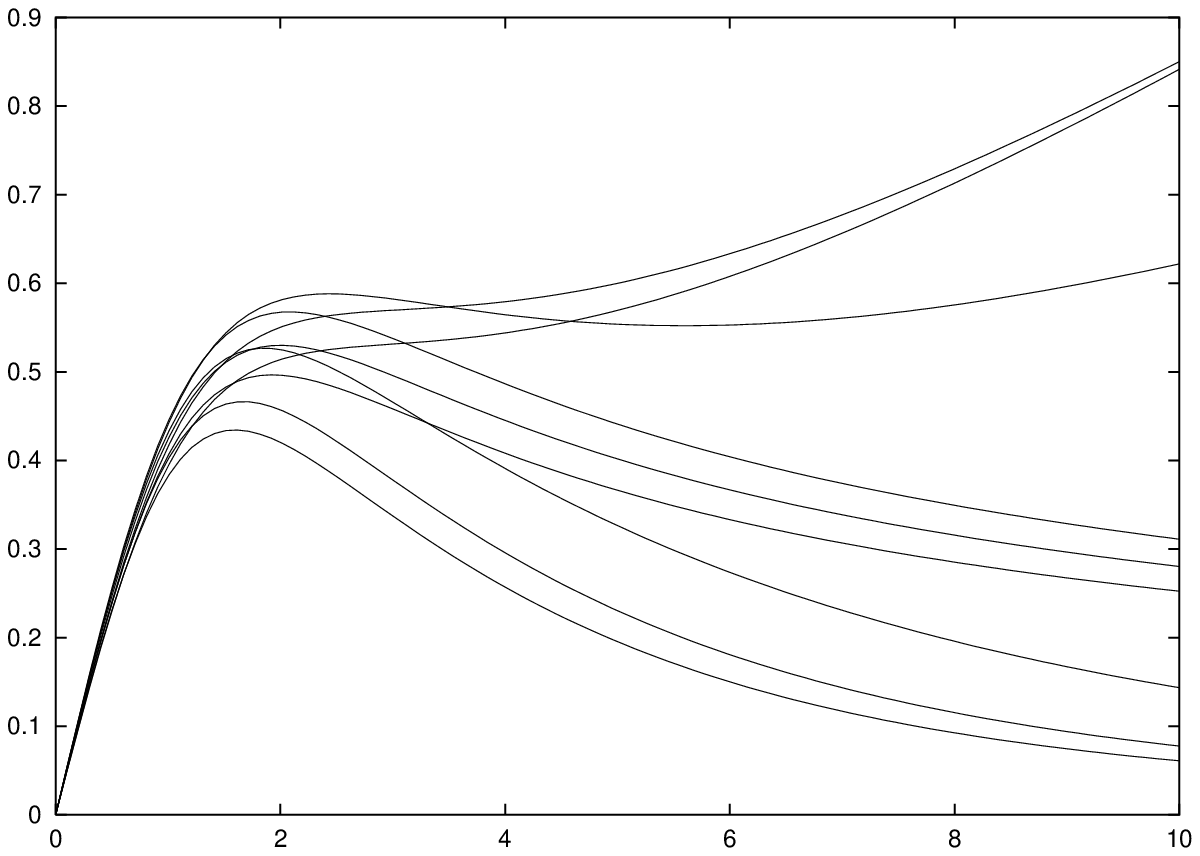} &  \epsfig{width=2.5in,file=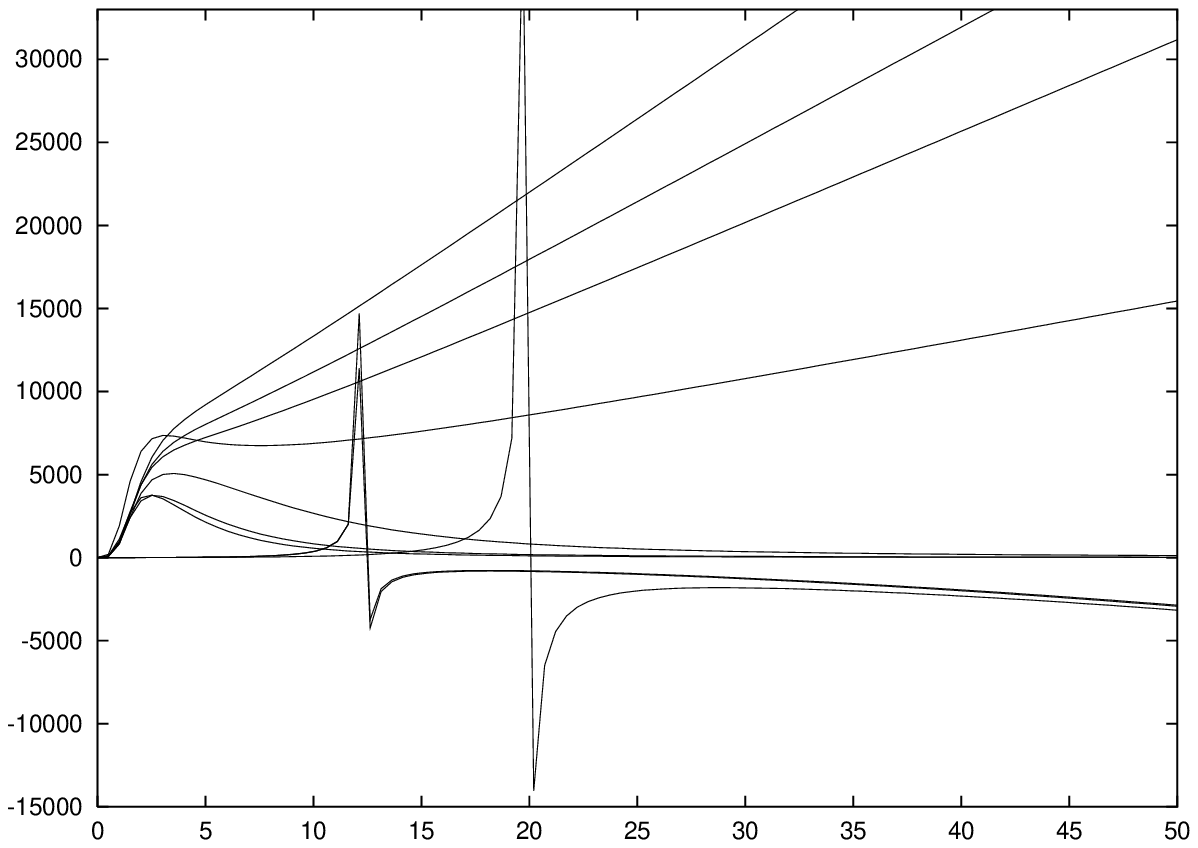} \\
 r  &  r \\
(a)  & (b)
\end{array}
$$	
\caption{ $(a)$ ${\rm U}_+$ (lower curves),  ${\rm U}_-$ (upper curves) for 
Chazy-Curzon-like  disks  with $\gamma=1$, $z_0=1.5$ and $p=q=0.5$,  $1.0$, and
$1.5$, and  $\rm{U}^2$  for $\gamma=1$, $z_0=1$ and $b=0.5$,  $1.0$, and
$1.5$. $(b)$ $h^2_+$ (scaled by a factor of 1000), $h^2_-$ (sharp curves )
for the same values of the parameters and  $h^2$ (upper curves scaled by a
factor of 1000) for    $\gamma=1$, $z_0=1.5$ and  $b=0.5$, $1.0$, $1.5$, and
$4$ (bottom curve).} \label{fig:cvm}
\end{figure*}

We now consider the  CRM for the same values of the parameters. All the
significant quantities can also be expresed in analytic form from the above
expressions but the results are so cumbersome that it is best just to analyze
them graphically. We first consider the two counterrotating streams circulating
along electro-geodesics.  In Fig.  \ref{fig:cvm}$(a)$ we plot the tangential
velocity curves of the counterrotating streams, ${\rm U}_+$, ${\rm U}_-$ for
disks  with $\gamma=1$, $z_0=1.5$ and $p=q=0.5$, $0.1$, and $1.5$, and
$\rm{U}^2$ for $\gamma=1$, $z_0=1.5$ and $b=0.5$, $0.1$, and $1.5$.  We can see
that ${\rm U}_+$  and ${\rm U}_-$ increase initially and then ${\rm U}_+$ 
falls to zero at infinity and always is less than the light velocity, whereas
${\rm U}_-$ increases monotonously. Therefore these disk models are well
behaved only  at the central regions. However, when the spacetime  is
electrostatic (or magnetostatic) one finds that the velocity  $\rm{U}^2$ for
these disks  is always less than the light velocity, but the disks with  
$b=0.5$ and $z_0 < 0.76$  cannot be built from the CRM  because $\rm{U}^2>1$
(not shown in the figure). One also finds  that  the inclusion   of electric
(magnetic) field and the increasing $z_0$ make less relativistic these disks.

In Fig. \ref{fig:cvm}$(b)$  we have drawn the specific angular momenta  of
counterrotating fluids $ h_+^2$, $ h_-^2$  for disks  with  $\gamma=1$,
$z_0=1.5$ and $p=q=0.5$, $0.1$, and $1.5$, and $h^2$ for $\gamma=1$, $z_0=1.5$,
$b=0.5$, $1.0$, $1.5$ and  $4$. We see that there is a strong change in the
slope of $ h_-^2$ at certain value of  $r$, which means that there is a strong
instability there. We also find regions with negative slope where the CRM is
also unstable. Also, $h_-^2$ presents instability after  certain value of r. 
Therefore these disks models are stable only at the central regions.  For
electrostatic (or magnetostatic) fields  the specific angular momenta  $h^2$ is
an increasing monotonous function of $r$ what corresponds to a stable CRM for
the disks. However, the CRM cannot be applied for $b=4.0$ (bottom curve). Thus 
the inclusion of electric (magnetic) field  can make unstable these disks
against radial perturbations. 

\begin{figure*}
$$
\begin{array}{cc}
 \epsilon _+, \  \epsilon _- , \ \epsilon _ \pm  & \sigma _- , \  - \sigma _+,
\   \ -  \sigma _{e \pm}, \  \mp  \sigma _{m \pm}   \\
\epsfig{width=2.5in,file=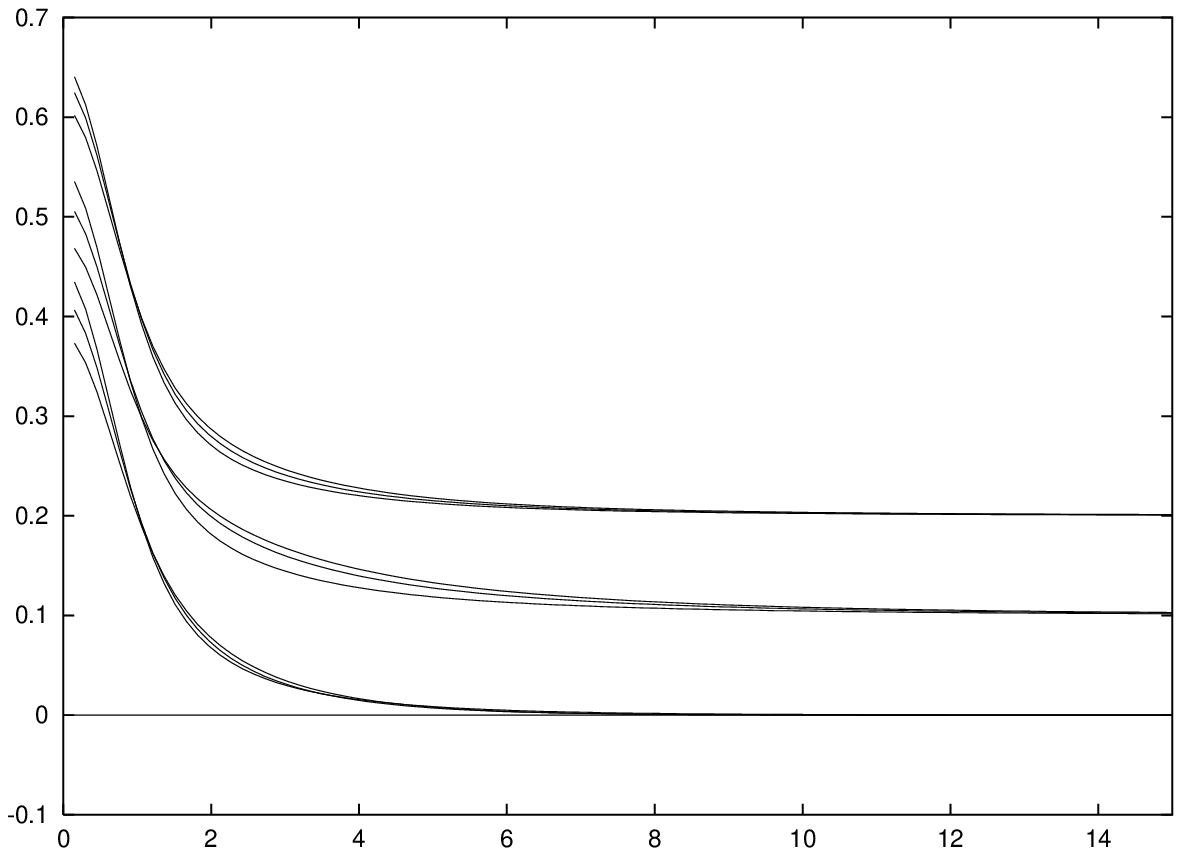} & \epsfig{width=2.5in,file=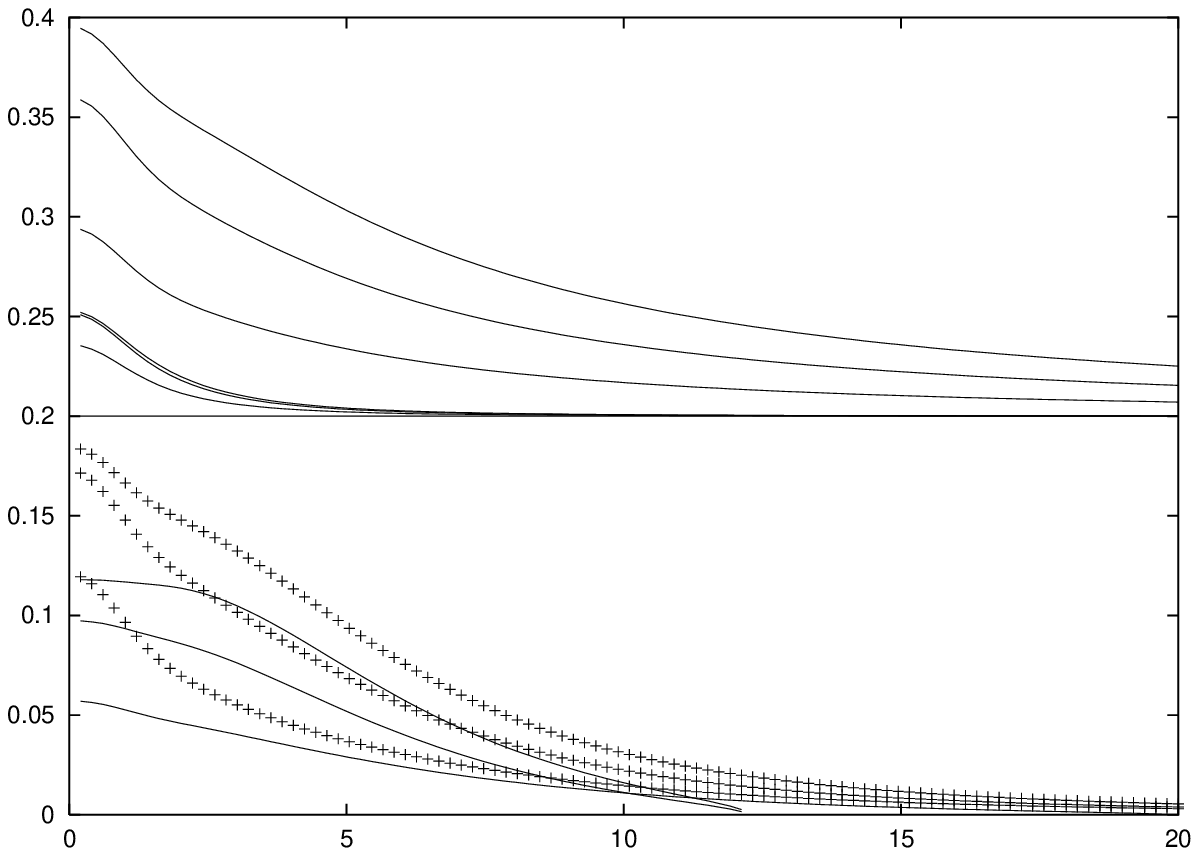}  \\
 r  &  r  \\
(a)  & (b)
\end{array}
$$	
\caption{$(a)$  $ \epsilon _-$ (lower curves), $ \epsilon _+$ (moved upwards a
factor of 0.1) for  Chazy-Curzon-like  disks  with $\gamma=1$, $z_0=1.5$ and
$p=q=0.5$ , $1.0$, and $1.5$ and $ \epsilon _\pm$  (upper curves moved upwards
a factor of 0.2) for  $\gamma=1$, $z_0=1.5$ and  $b=0.5$, $1.0$ and  $1.5$.
$(b)$ $\sigma _-$,  $\sigma _+$ (curves with crosses),
$\sigma _{e \pm}$ (moved upwards a factor of 0.2) and $\sigma _{m \pm}$ (upper
curves moved also upwards a factor of 0.2) for the same values of the
parameters.} \label{fig:cecrc}
\end{figure*}

In Fig. \ref{fig:cecrc}$(a)$ the plots of the  mass densities  $\epsilon _+$,  
$\epsilon _-$  are shown for disks  with $\gamma=1$, $z_0=1.5$ and $p=q=0.5$,
$0.1$, and $1.5$, and $\epsilon _\pm$ for $\gamma=1$, $z_0=1.5$, $b=0.5$,
$1.0$, and $1.5$, as functions of $r$. The mass density $ \epsilon _+$ is
always positive, falling to zero at infinity, whereas $ \epsilon _-$ becomes
negative after  some value of $r$ and then falls to zero at infinity. The  mass
densities  $\epsilon _\pm$ for electrostatic (or magnetostatic) fields  are
positive everywhere on the disks, falling to zero at infinity. We also see
that  the presence of electromagnetic  field decreases the  progradee and
retrogade  mass densities   at the central region of the disks and  later
increases them.

In Fig. \ref{fig:cecrc}$(b)$ the plots of the electric charge densities   $
\sigma _+$,    $\sigma _-$  are shown  for disks  with $\gamma=1$, $z_0=1.5$
and $p=q=0.5$, $0.1$, and $1.5$, and $\sigma _{e \pm}$ and  $\sigma _{m \pm}$
for $\gamma=1$, $z_0=1.5$, $b=0.5$, $1.0$, and $1.5$, as functions of $r$. We
see that $\sigma _+$ falls to zero at infinity, whereas  $\sigma _-$ becomes
later imaginary.   $\sigma _{e \pm}$ and  $\sigma _{m \pm}$  present a maximun
at the disks center, then decrease monotonously, falling to zero at infinity,
and are always  real quantities.

\begin{figure*}
$$
\begin{array}{c}
 - \sigma _+, \  \sigma _-     \\
\epsfig{width=2.5in,file=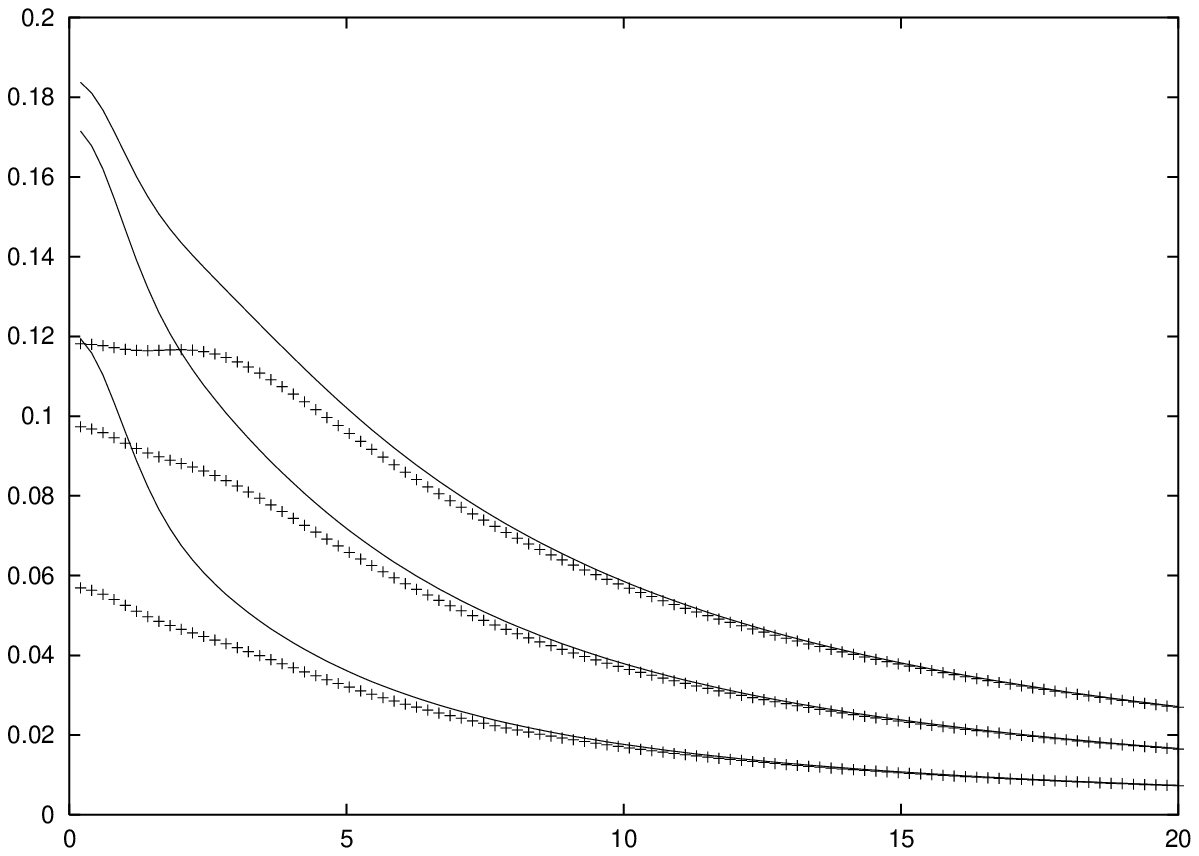}  \\
r  
\end{array}
$$	
\caption{$\sigma _+$ (solid curves) and   $\sigma _-$  for not
electro-geodesic  Chazy-Curzon-like  disks  with $\gamma=1$, $z_0=1.5$ and
$p=q=0.5$ (lower curves), $1.0$, and $1.5$ (upper curves).} \label{fig:ccng}
\end{figure*}

As with the electro-geodesic counterrotating fluids we obtain expressions that
becomes imaginary in portions of the disks, we also consider non
electro-geodesics CRM for these disks. When the two counterrotating streams do
not move on electro-geodesics but  have  equal and opposite tangential
velocities the graphics for the  velocities, energy  densities and specific
angular momenta  of both streams will  have the same behavior that in  the
electrostatic (magnetostatic) case. The expressions for  the electric charge
densities  are differents and  are plotted in Fig. \ref{fig:ccng}, being always
real quantities. Thus, for  Chazy-Curzon-like fields we can build 
electro-geodesic counterrotating thin disk sources with a well behaved central
region,  whereas when we have counterrotating streams not moving along
electro-geodesic or  electrostatic (magnetostatic) fields we obtain  physically
acceptable counterrotating disk for many values of the parameters. 

\subsection{CRM for Zipoy-Voorhees-like disks}

The second  family of  solutions considered  is a Zipoy-Voorhees-like solution
which can be written as 
\begin{subequations}\begin{eqnarray}
e^\nu    &=& \frac{2 (x^2 - 1)^{\gamma/2}}{(1 + a)(x + 1)^\gamma + (1 - a)(x -
1)^\gamma} , \\
        & & \nonumber   \\
\lambda &=&\frac{\gamma ^2}{2} \ln \left [ \frac {x^2-1}{x^2-y^2}  \right ], \\
	& & \nonumber   \\
A_t &=& \frac { \sqrt 2 p [(x + 1)^\gamma - (x - 1)^\gamma] }{(1 + a)(x +
1)^\gamma +(1 - a)(x - 1)^\gamma}, \\
        & & \nonumber   \\
A_\varphi & = &\sqrt 2 k\gamma qy,
\end{eqnarray}\label{eq:zv}\end{subequations}
where $a^2=1+b^2$, with  $b^2 = p^2 + q^2 $, and $\gamma$ is a real constant.
Here $p$ and $q$ are again the  electric and magnetic parameters, respectively.
$x$ and $y$ are the prolate spheroidal coordinates, related to the Weyl
coordinates by 
\begin{equation}
r^2   = k^2 (x^2-1)(1-y^2),  \quad  \quad z +z_0  = kxy. \label{eq:coorp}
\end{equation}

This solution can also be generated, in these coordinates,  using the
well-known  complex potencial formalism proposed by Ernst \cite{E2} from the  
Zipoy-Voorhees vacuum solution \cite{Z,V}, also known as the Weyl
$\gamma$-solution \cite{W1,W2}, by choosing the parameter $q$ of Ref. \cite{
E2} as complex.  For  $q=0$ we also have an electrostatic solution and for
$p=0$  one  obtains its  magnetostatic equivalent \cite{B1}. The case $q=0$ and
$\gamma =1$ corresponds to the   Reissner-Nordstr\"{o}m  solution \cite{KSHM},
in which case $a= m/k$, $p=e/k$, with $k^2=m^2-e^2$, so that $a^2=1+p^2$,  $m$
and $e$ being the mass and  charge parameters, respectively. When    $p=0$ and
$\gamma =2$ one arrives to  a Taub-NUT-like  magnetostatic solution.   This  
solution can also   be generated, in these coordinates,  using a well-know
theorem  proposed by Bonnor (see Ref. \cite{B2}) from  the Taub-NUT  vacuum
solution. Note that for $b=0$, it reduces to the Darmois metric \cite{KSHM}.

From the above expressions we can compute the physical quantities associated
with the disks. We obtain

\begin{subequations}\begin{eqnarray}
\tilde {\epsilon} &=&  \frac{8 \gamma \bar{y}(\bar{x}^2 - 1)^{\frac{\gamma}{2}
(1 - \gamma)} (\bar{x}^2 - \bar{y}^2)^{\frac{1}{2}(\gamma^2 -4)}} {[(1 +
a)(\bar{x} + 1)^\gamma + (1 - a)(\bar{x} - 1)^\gamma]^2} \nonumber \\
&& \nonumber	\\
&& \times \left\{ \bar{x} [(1 + a)(\bar{x} + 1)^\gamma  + (1 - a)(\bar{x} -
1)^\gamma] [\bar{x}^2 + (\gamma - 1)\bar{y}^2   - \gamma] \right . \\
& & 	\nonumber	\\
&& \left . \quad - (\bar{x}^2 - 1)(\bar{x}^2 - \bar{y}^2)[(1 + a)(\bar{x} +
1)^{\gamma-1} + (1 - a)(\bar{x} - 1)^{\gamma - 1}] \right\} , \nonumber \\
&&	\nonumber	\\
\tilde {p}_\varphi &=&  \frac {8 \gamma ^2 \bar{x} \bar{y} (1 - \bar{y}^2)
(\bar{x}^2 - 1)^{\frac{1}{2} \gamma (1 - \gamma)} (\bar{x}^2 -
\bar{y}^2)^{\frac 12  (\gamma^2 - 4)}}{[(1 + a)(\bar{x} + 1)^\gamma + (1 -
a)(\bar{x} - 1)^\gamma]}, \\
&&	\nonumber	\\
\tilde{\rm j}_t &=& \frac {16 \sqrt 2 \gamma p \bar{y} (\bar{x}^2 - 1)^{\frac
12 \gamma (3 - \gamma)} (\bar{x}^2 - \bar{y}^2)^{\frac 12  (\gamma^2 - 2)}}{[(1
+ a)(\bar{x} + 1)^\gamma + (1 - a)(\bar{x} - 1)^\gamma]^3}, \\
&&	\nonumber	\\
{\rm j}_\varphi &=& -  \frac {4 \sqrt 2 \gamma q \bar{x} (1 - \bar{y}^2)
(\bar{x}^2 - 1)^{\frac 12 \gamma (1 - \gamma)} (\bar{x}^2 - \bar{y}^2)^{\frac
12  (\gamma^2 - 2)}}{[(1 + a)(\bar{x} + 1)^\gamma +(1 - a)(\bar{x} -
1)^\gamma]},
\end{eqnarray}\end{subequations}

where $\tilde {\epsilon} = k \epsilon$, $\tilde {p}_\varphi = k p_\varphi$ and
$\tilde{\rm j}_t = k {\rm j}_t$. $\bar{x}$ and  $\bar{y}$ are given by
\begin{subequations}\begin{eqnarray}
2\bar{x}  & = \sqrt {\tilde{r}^2 + (\alpha + 1)^2} + \sqrt {\tilde{r}^2 +
(\alpha - 1)^2}, \label{eq:xbar} \\
     &  \nonumber       \\
2\bar{y}  & = \sqrt {\tilde{r}^2 + (\alpha + 1)^2} - \sqrt {\tilde{r}^2 +
(\alpha - 1)^2},\label{eq:ybar}
\end{eqnarray}\end{subequations}
where $\tilde{r}=r/k$ and $\alpha = z_0/k$, with $\alpha >1$.

\begin{figure*}
$$
\begin{array}{cc}
\tilde \epsilon, \  \tilde p_\varphi  & \tilde {\rm j}_t, \ {\rm j}_\varphi \\
\epsfig{width=2.5in,file=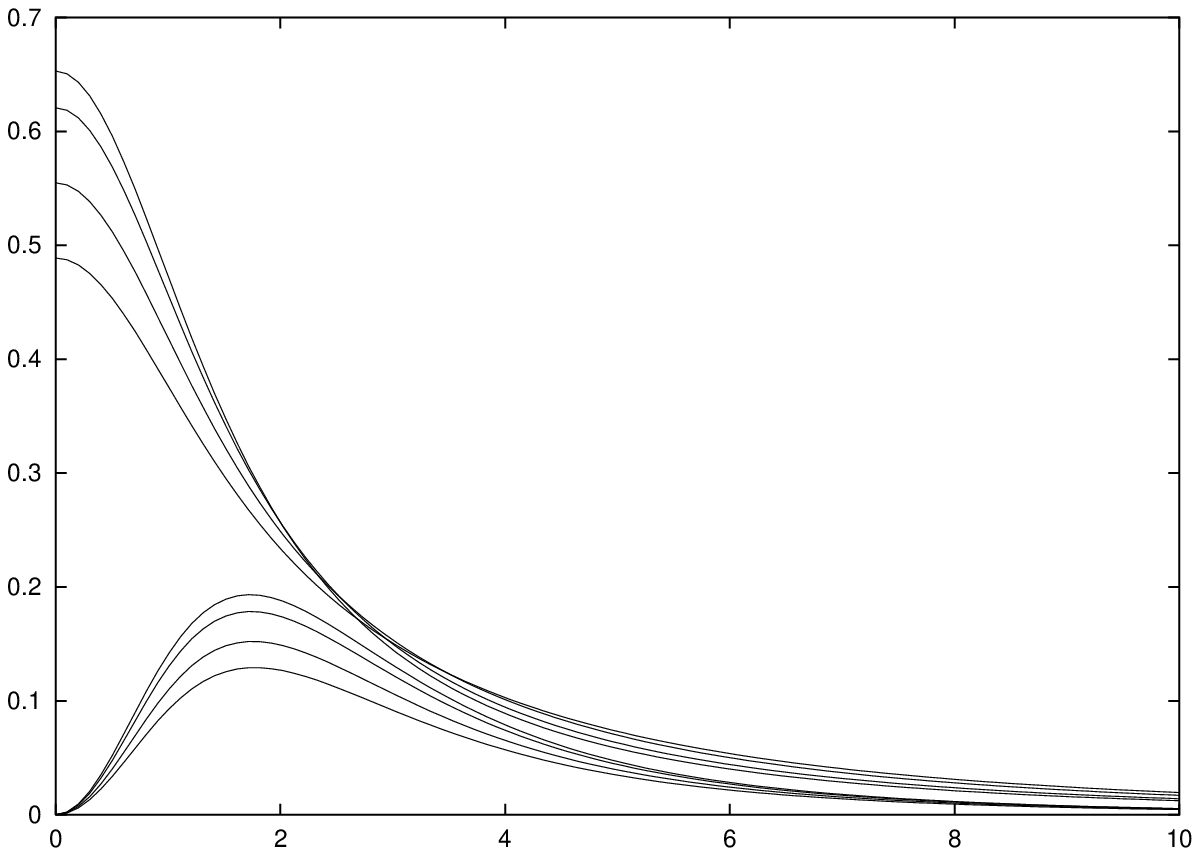} & \epsfig{width=2.5in,file=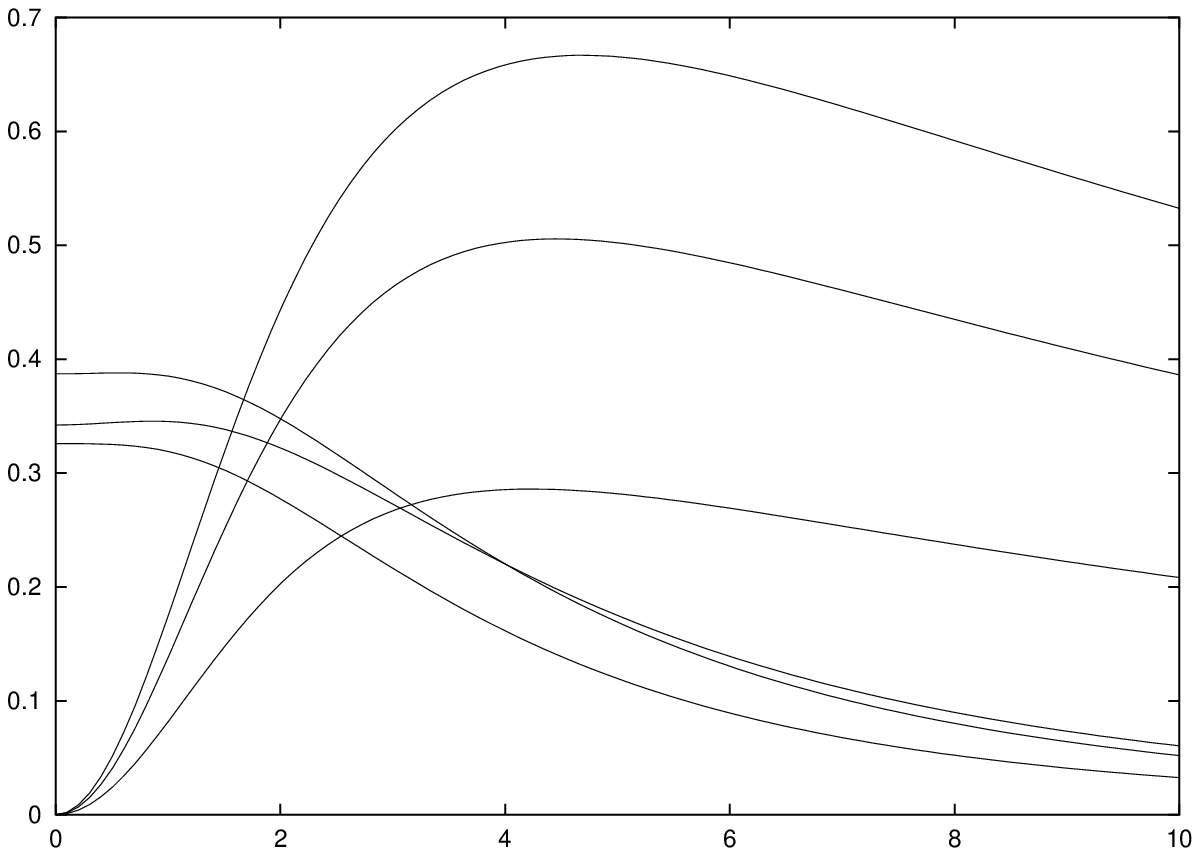} \\
\tilde r & \tilde r    \\
(a)     &   (b)
\end{array}
$$	
\caption{$(a)$ $\tilde \epsilon$ (upper curves ) and $\tilde p_\varphi$ (scaled
by a factor of 2),  as functions of $\tilde r$ for  Zipoy-Voorhees-like  disks 
with $\gamma=2$, $\alpha=2.5$ and $p=q=0$ (top curves), $0.5$, $1.0$, and $1.5$
(bottom curves).  $(b)$ $\tilde {\rm j}_t$ (lower curves  scaled by a factor of
10) and ${\rm j}_\varphi$ for the same values  of the parameters.}
\label{fig:zepco}
\end{figure*}

We consider as an example a particular case, when  $\gamma =2$. We also study
Zipoy-Voorhees-like disks for other values of $\gamma$, but in all cases we
found a similar behavior. In Fig.  \ref{fig:zepco} the plots of the quantities 
$\tilde {\epsilon}$, $\tilde {p}_\varphi$, $\tilde{\rm j}_t$ and ${\rm
j}_\varphi$ are presented for disks with $\alpha=2.5$ and $p=q=0$, $0.5$,
$1.0$, and $1.5$, as functions of $\tilde{r}$. We see that these functions have
a similar  behavior to the previous case. Equally, the  relevant  quantities of
the CRM are shown in following figures  for the same values of the parameters,
also as functions of $\tilde{r}$. We also first consider the two
counterrotating streams circulating along electro-geodesics. 

\begin{figure*}
$$
\begin{array}{cc}
{\rm U}_+, \ \ -{\rm U}_-, \ \   \rm U^2 &  \tilde h^2_+, \ \  \tilde h^2_-,
\tilde h^2   \\
\epsfig{width=2.5in,file=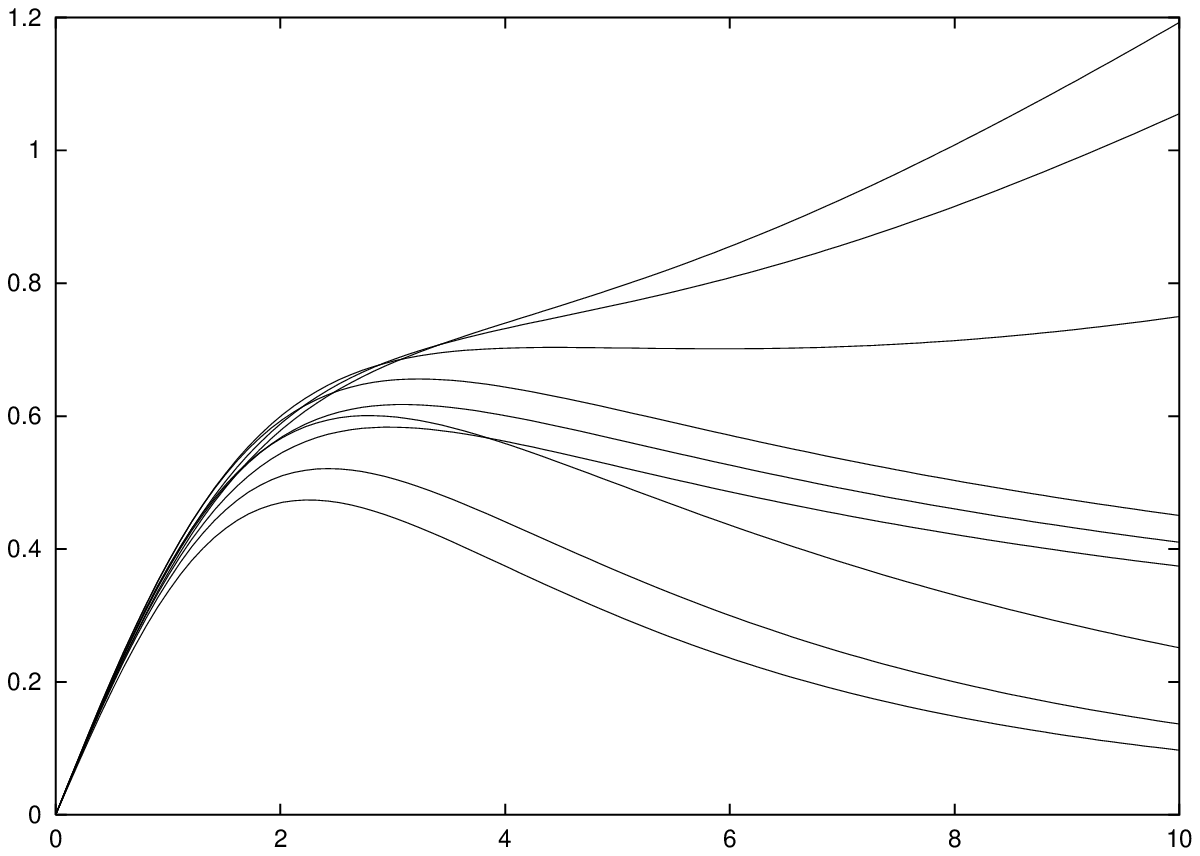} &  \epsfig{width=2.5in,file=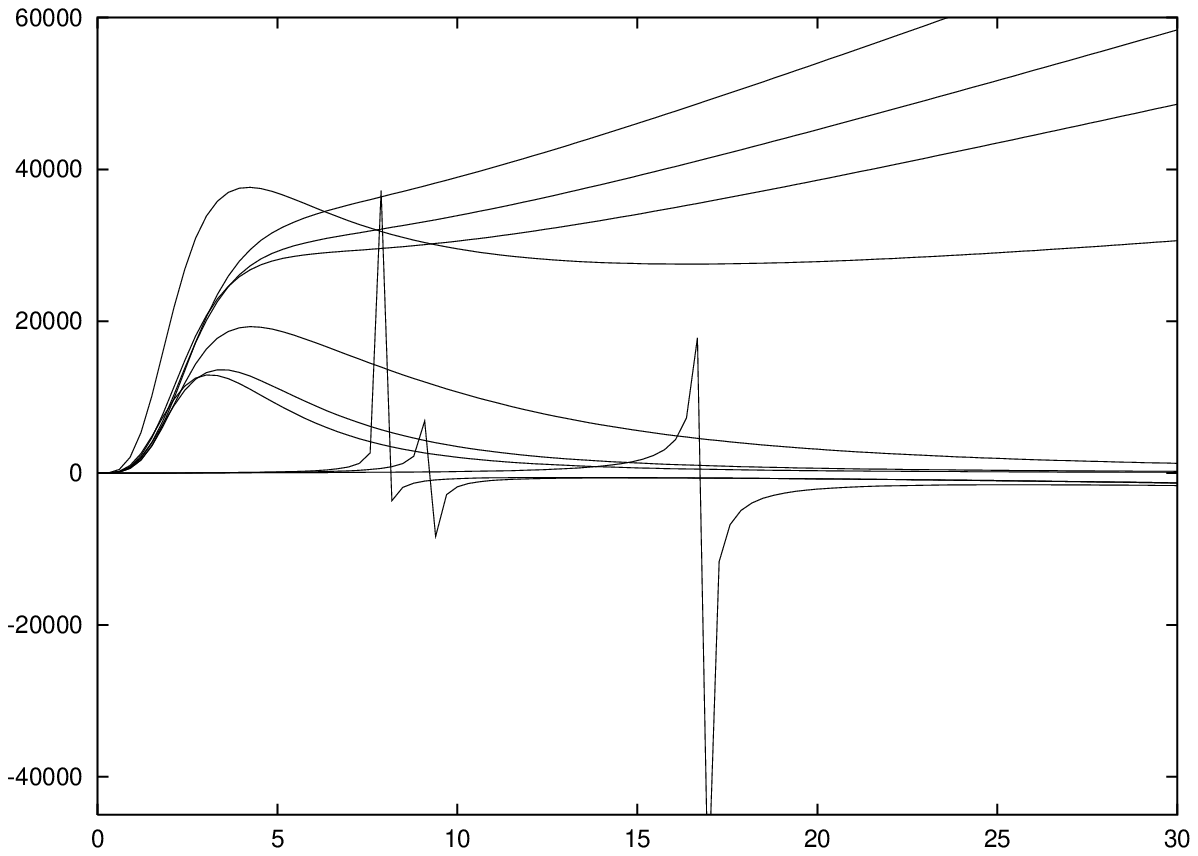} \\
 \tilde r  &  \tilde r \\
(a)  & (b)
\end{array}
$$	
\caption{ $(a)$ ${\rm U}_+$ (lower curves),  ${\rm U}_-$ (upper curves) as
functions of $\tilde r$ for  Zipoy-Voorhees-like  disks  with $\gamma=2$,
$\alpha=2.5$ and $p=q=0.5$, $1.0$, and $1.5$  and  $\rm{U}^2$  for $\gamma=2$,
$\alpha=2.5$ and $b=0.5$, $1.0$, and $1.5$ . $(b)$ $\tilde h^2_+$ (scaled by a
factor of 1000),   $\tilde h^2_-$ (sharp curves) for the same values of the
parameters and $\tilde h^2$ (upper curves scaled by a factor of 1000) for 
$\gamma=2$, $\alpha=2.5$ and  $b=0.5$, $1.0$, $1.5$ and $4.0$ (bottom curve). 
}  \label{fig:zvm}
\end{figure*}

Here  the tangential velocity ${\rm U}_+$  (Fig. \ref{fig:zvm}$(a)$) is also
always  less than the light velocity, falling to zero at infinity, whereas
${\rm U}_-$ is an increasing monotonously function. Thus these disk models  are
also just  well behaved  at the central regions. In electrostatic
(magnetostatic) case,  one also  finds that the velocity  $\rm{U}^2$ for these
disks  is always less than the light velocity, but  the disks with $b=0.5$ and
$\alpha < 1.58 $   cannot be constructed from the CRM because $\rm{U}^2>1$. One
also finds  that  the inclusion   of electric (magnetic) field and the
increasing $\alpha$ make  less  relativistic these disks. Moreover, we  see
that $\tilde h_+^2$  and $\tilde h_-^2$  (Fig. \ref{fig:zvm}$(b)$) present
instabilities in the same way than the precedent case, and  that for
electrostatic (or magnetostaic) disks (upper curves) the presence of electric
(or magnetic) field  can make unstable the CRM against radial perturbations.
Thus the CRM cannot apply for $b=4$ (bottom curve).

\begin{figure*}
$$
\begin{array}{cc}
\tilde \epsilon _-, \ \tilde \epsilon _+,  \ \tilde \epsilon _ \pm  & \tilde
\sigma _- , \  - \tilde  \sigma _+,   \ -  \tilde \sigma _{e \pm}, \  \mp 
\tilde \sigma _{m \pm}   \\
\epsfig{width=2.5in,file=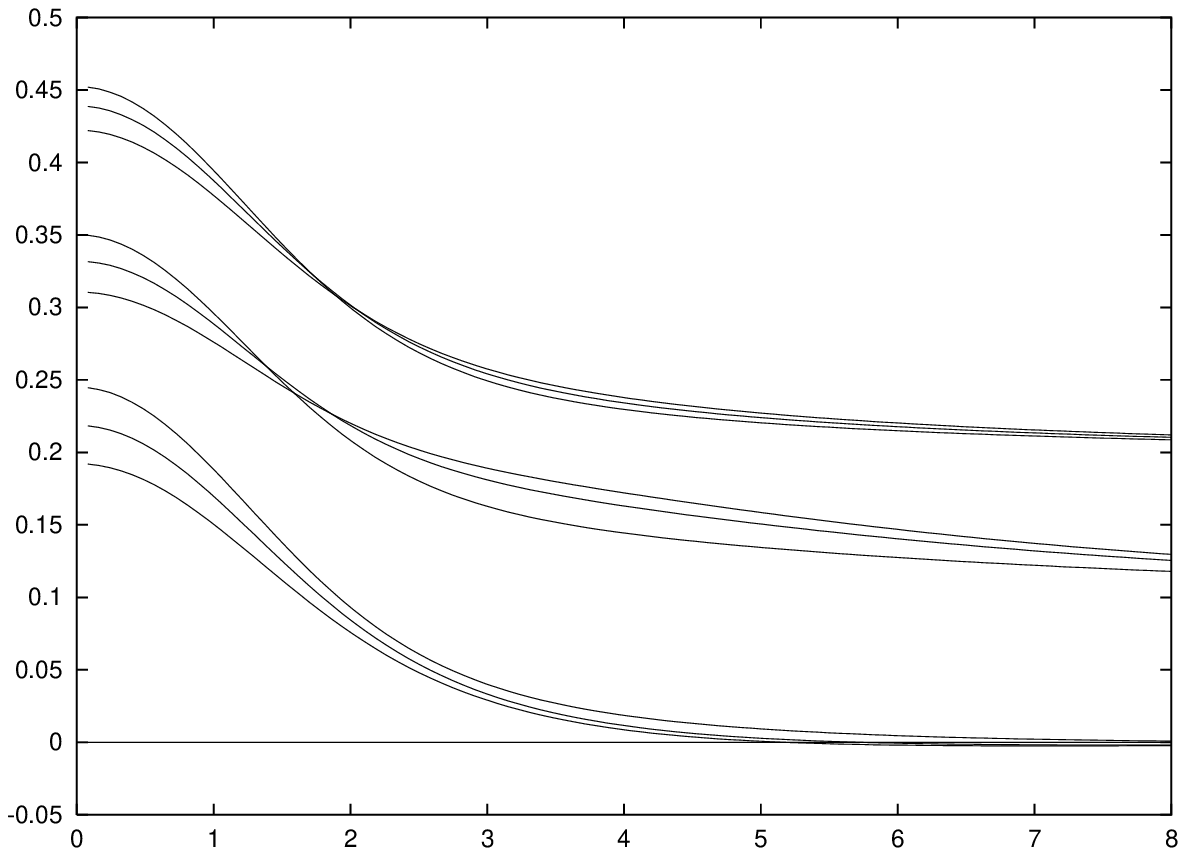} & \epsfig{width=2.5in,file=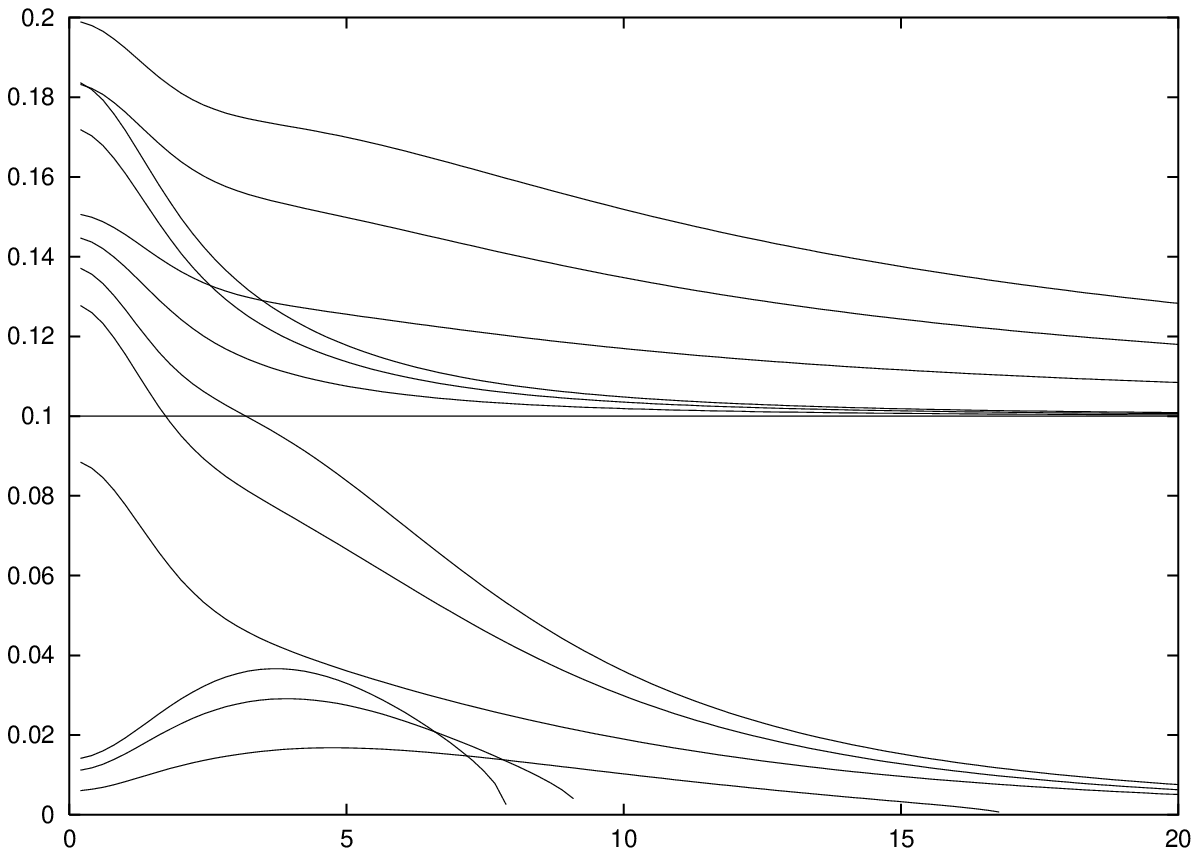}  \\
\tilde r  & \tilde r \\
(a)  & (b)
\end{array}
$$	
\caption{$(a)$ $\tilde \epsilon _-$ (lower curves), $\tilde \epsilon _+$ (moved
upwards a factor of 0.1)  for   Zipoy-Voorhees-like  disks  with $\gamma=2$,
$\alpha=2.5$ and $p=q=0.5$ , $1.0$, and $1.5$ and $\tilde \epsilon _\pm$ 
(upper curves moved upwards a factor of 0.2) for  $\gamma=2$, $\alpha=2.5$ and 
$b=0.5$, $1.0$ and  $1.5$.  $(b)$  $\tilde \sigma _-$ (lower curves),  $\tilde
\sigma _+$,  $\tilde \sigma _{e \pm}$  (moved upwards a factor of 0.1) and  
$\tilde \sigma _{m \pm}$ (upper curves moved also upwards a factor of 0.1 ) for
the same values of the parameters.} \label{fig:zecrc}
\end{figure*}

The mass density $\tilde \epsilon _+$ (Fig. \ref{fig:zecrc}$(a)$) is also
always positive, falling to zero at infinity, whereas $\tilde \epsilon _-$
becomes negative after  some value of $\tilde{r}$ and then falls to zero at
infinity. We also see that  the presence of electromagnetic  field decreases
the  progradee and retrogade  mass densities   at the central region of the
disks and  later increases them. For electrostatic (or magnetostatic) fields 
$\tilde \epsilon _\pm$  is positive everywhere on the disks, falling to zero at
infinity. We also  find that $\tilde \sigma _+$   (Fig. \ref{fig:zecrc}$(b)$) 
falls to zero at infinity, whereas  $\tilde \sigma _-$ becomes later imaginary.
The   electrostatic and magnetostatic charge  densities, $\tilde \sigma _{e
\pm}$  and $\tilde \sigma _{m \pm}$, are always   real quantities, falling to
zero at infinity.

\begin{figure*}
$$
\begin{array}{c}
- \tilde \sigma _+, \  \tilde \sigma _-     \\
\epsfig{width=2.5in,file=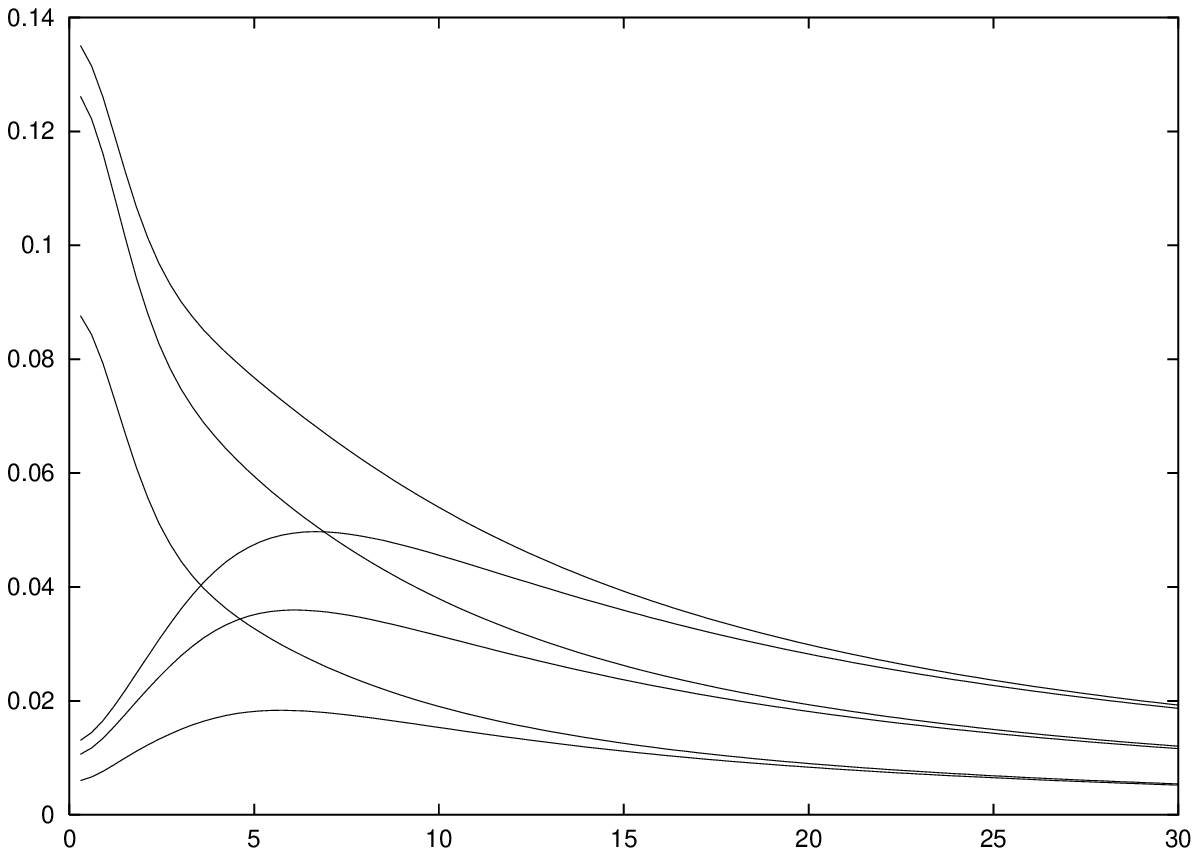}  \\
\tilde  r  
\end{array}
$$	
\caption{$\tilde \sigma _+$ (upper curves)  and   $\tilde \sigma _-$  for not
electro-geodesic  Zipoy-Voorhees-like   disks  with $\gamma=2$, $\alpha=2.5$
and $p=q=0.5$ (lower curves), $1.0$, and $1.5$ (upper curves).}
\label{fig:zcng}
\end{figure*}

Then,  when we consider  the two counterrotating streams  not moving on
electro-geodesics but  with   equal and opposite tangential velocities, one
also finds that  quantities ${\rm U}^2$, $\tilde \epsilon _\pm$, and $\tilde
h^2$  have the same behavior than  in the electrostatic (magnetostatic) case. 
The electric charge densities is shown   in Fig. \ref{fig:zcng}, being always
real quantities. Therefore,  for Zipoy-Voorhees-like  fields we can build 
electro-geodesic counterrotating thin disk sources  with only a  well behaved
central region,  whereas for  counterrotating streams not circulating along
electro-geodesic or for electrostatic (magnetostatic) fields,  we obtain disks
with a physically acceptable CRM for many values of the parameters. 

\subsection{CRM for Bonnor-Sackfield-like disks}

The third family of solutions considered is a  Bonnor-Sackfield-like solution
which is given by
\begin{subequations}\begin{eqnarray}
e^\nu &=& \frac {2}{(1 + a) e^{\gamma \cot ^{-1}u} + (1 - a) e^{-\gamma \cot
^{-1}u}} , \\
&&        \nonumber   \\
\lambda &=&  - \frac{\gamma^2}{2} \ln \left[ \frac{u^2 + 1}{u^2 + v^2} \right] 
, \\
&&         \nonumber   \\
A_t &=& \frac{\sqrt{2} p [e^{\gamma \cot ^{-1} u} - e^{-\gamma \cot ^{-1}
u}]}{(1 + a) e^{\gamma \cot ^{-1} u} + (1 - a) e^{-\gamma \cot ^{-1} u}}, \\
&& \nonumber   \\
A_\varphi &=& \sqrt 2 k \gamma q v,
\end{eqnarray}\label{eq:bs}\end{subequations}
where $a^2=1+b^2$, with  $b^2 = p^2 + q^2 $, and $\gamma$ is a real constant.
Here $p$ and $q$ are again the  electric and magnetic parameters, respectively.
$u$ and $v$ are the oblate spheroidal coordinates, related to the Weyl
coordinates by 
\begin{equation}
r^2   = k^2 (u^2+1)(1-v^2),  \quad  \quad z +z_0  = kuv.
\end{equation}
As in the precedent cases this solution can  be generated, in these
coordinates, following  Ernst's method \cite{E2} from the   Bonnor-Sackfield 
vacuum solution \cite{BS}.

From the above expressions we can compute the physical quantities associated
with the disks. We obtain

\begin{subequations}\begin{eqnarray}
\tilde \epsilon &=& \frac{8\gamma \bar v(\bar u ^2 + 1
)^{\frac{\gamma^2}{2}}}{(\bar u + \bar v)^{\frac{\gamma^2}{2} + 2}[(1 + a)
e^{\gamma \cot^{-1}\bar u} +  (1 - a) e^{-\gamma \cot^{-1}\bar u}]^2} \nonumber
\\
&&	\\
&&\times \left\{[{\bar u}^2 + {\bar v}^2 -\gamma \bar u(1 - {\bar v}^2)](1 + a)
e^{\gamma \cot^{-1}\bar u} -[\bar u^2 + \bar v ^2 + \gamma \bar u(1 - \bar
v^2)] (1 - a) e^{-\gamma \cot^{-1}\bar u} \right\} , \nonumber \\
&&	\nonumber	\\
\tilde p _\varphi &=&  \frac { 8\gamma^2 \bar u \bar v (1 - \bar v ^2)(\bar u +
1)^{\frac{\gamma^2}{2}}}{ (\bar u + \bar v)^{\frac{\gamma^2}{2} + 2}[(1 + a)
e^{\gamma \cot^{-1}\bar u} + (1 - a) e^{-\gamma \cot^{-1}\bar u}] }   , \\
&&	\nonumber	\\
\tilde {\rm j}_t &=&  \frac { 16 \sqrt 2  \gamma p \bar v (\bar u +
1)^{\frac{\gamma ^2}{2}} }{ (\bar u + \bar v)^{\frac{\gamma^2}{2} + 1}[(1 + a)
e^{\gamma \cot^{-1}\bar u} + (1 - a) e^{-\gamma \cot^{-1}\bar u}]^3 }   , \\
&&	\nonumber	\\
{\rm j}_\varphi &=& - \frac{ 4 \sqrt 2  \gamma q \bar u (1 - \bar v^2) (\bar u
+ 1)^{\gamma ^2/2}}{ (\bar u + \bar v)^{\frac{\gamma^2}{2} + 1}[(1 + a)
e^{\gamma \cot^{-1}\bar u} +  (1 - a) e^{-\gamma \cot^{-1}\bar u}] }   , 
\end{eqnarray}\end{subequations}
where $\tilde {\epsilon} = k \epsilon$, $\tilde {p}_\varphi = k p_\varphi$ and
$\tilde{\rm j}_t = k {\rm j}_t$. $\bar u$ and  $\bar v$ are given by
\begin{subequations}\begin{eqnarray}
\sqrt 2 \bar u &=& \sqrt{ [(\tilde r ^2 + \alpha ^2 -1)^2 + 4\alpha^2]^{1/2} +
(\tilde r ^2 + \alpha ^2 - 1)} , \\
               & & \nonumber   \\
\sqrt 2 \bar v &=& \sqrt{ [(\tilde r ^2 + \alpha ^2 -1)^2 + 4\alpha^2]^{1/2} -
(\tilde r ^2 + \alpha ^2 - 1)} ,
\end{eqnarray}\end{subequations}
where $\tilde{r}=r/k$ and $\alpha = z_0/k$.

\begin{figure*}
$$
\begin{array}{cc}
\tilde \epsilon, \  \tilde p_\varphi  & \tilde {\rm j}_t, \ {\rm j}_\varphi \\
\epsfig{width=2.5in,file=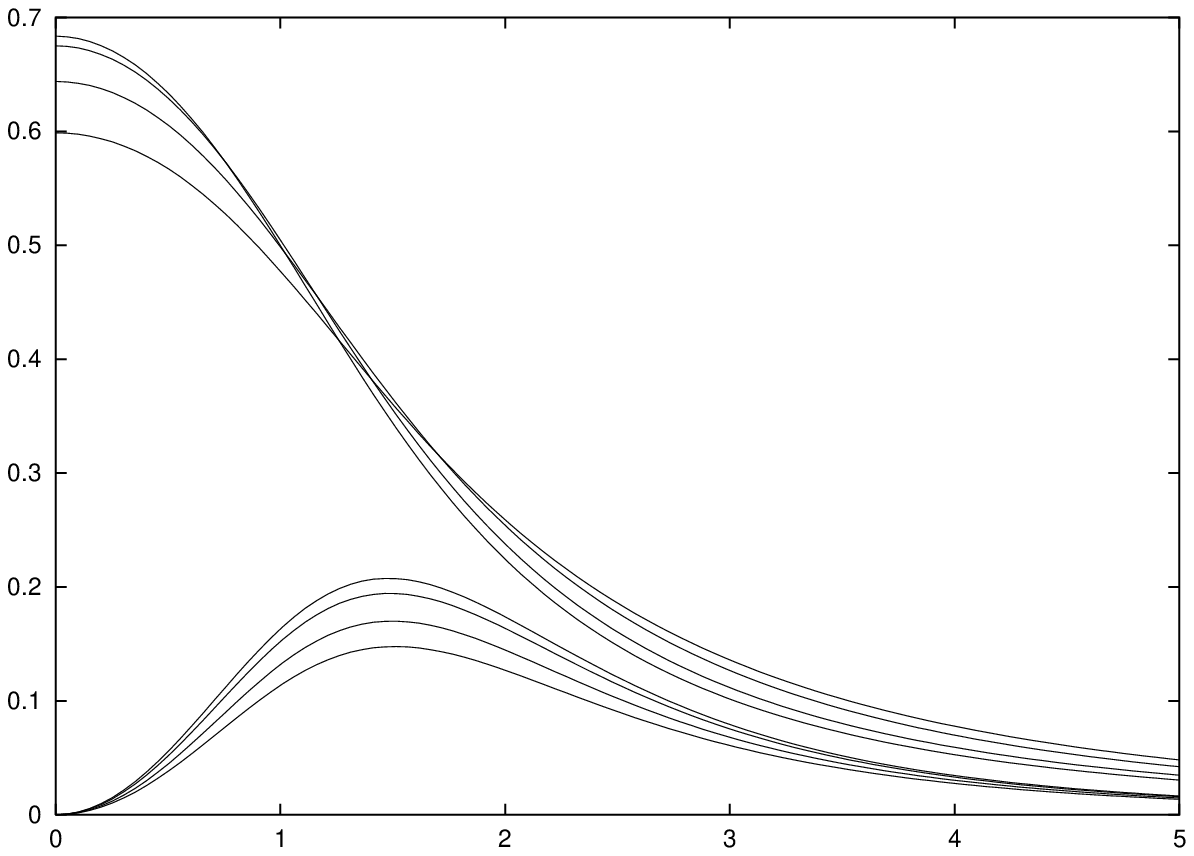} & \epsfig{width=2.5in,file=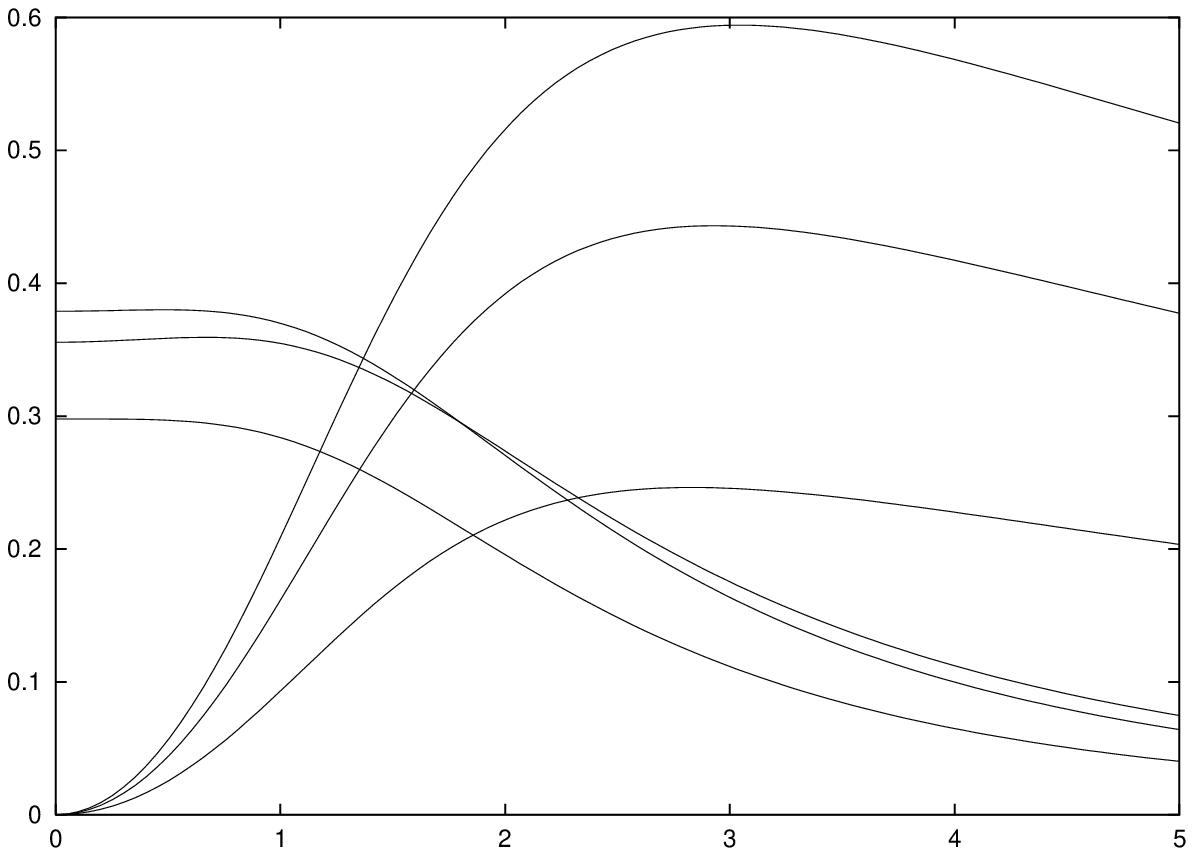} \\
\tilde r & \tilde r    \\
(a)     &   (b)
\end{array}
$$	
\caption{$(a)$ $\tilde \epsilon$ (upper curves ) and $\tilde p_\varphi$ (scaled
 by a factor of 2.5),  as functions of $\tilde r$ for  Bonnor-Sackfield-like
 disks  with $\gamma=1$, $\alpha=1.5$ and $p=q=0$ (top curves), $0.5$, $1.0$,
 and $1.5$ (bottom curves).  $(b)$ $\tilde {\rm j}_t$ (lower curves scaled by a
 factor of 5) and ${\rm j}_\varphi$ for the same values  of the
 parameters.}\label{fig:bepco}
\end{figure*}

\begin{figure*}
$$
\begin{array}{cc}
{\rm U}_+, \ \ -{\rm U}_-, \ \   \rm U^2 & \tilde h^2_+, \ \  \tilde h^2_-, \ \
\tilde h^2   \\
\epsfig{width=2.5in,file=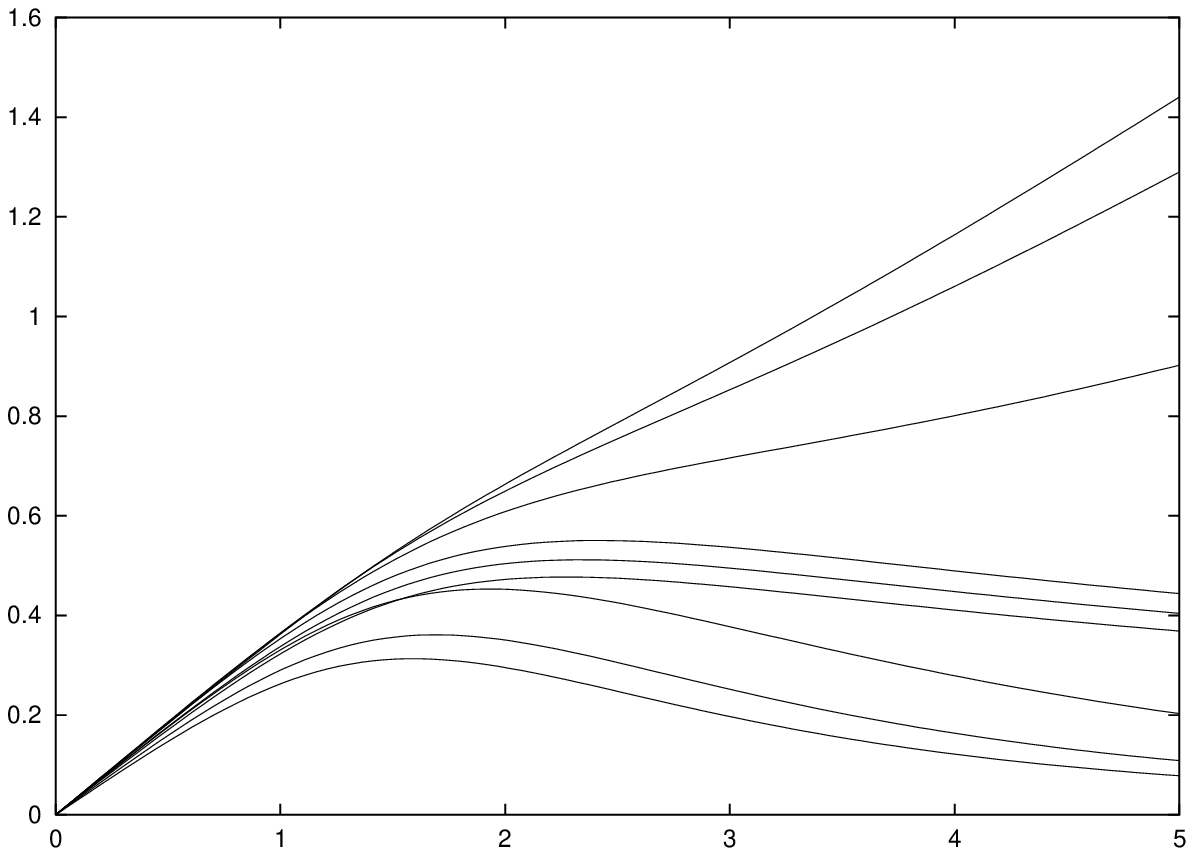} &  \epsfig{width=2.5in,file=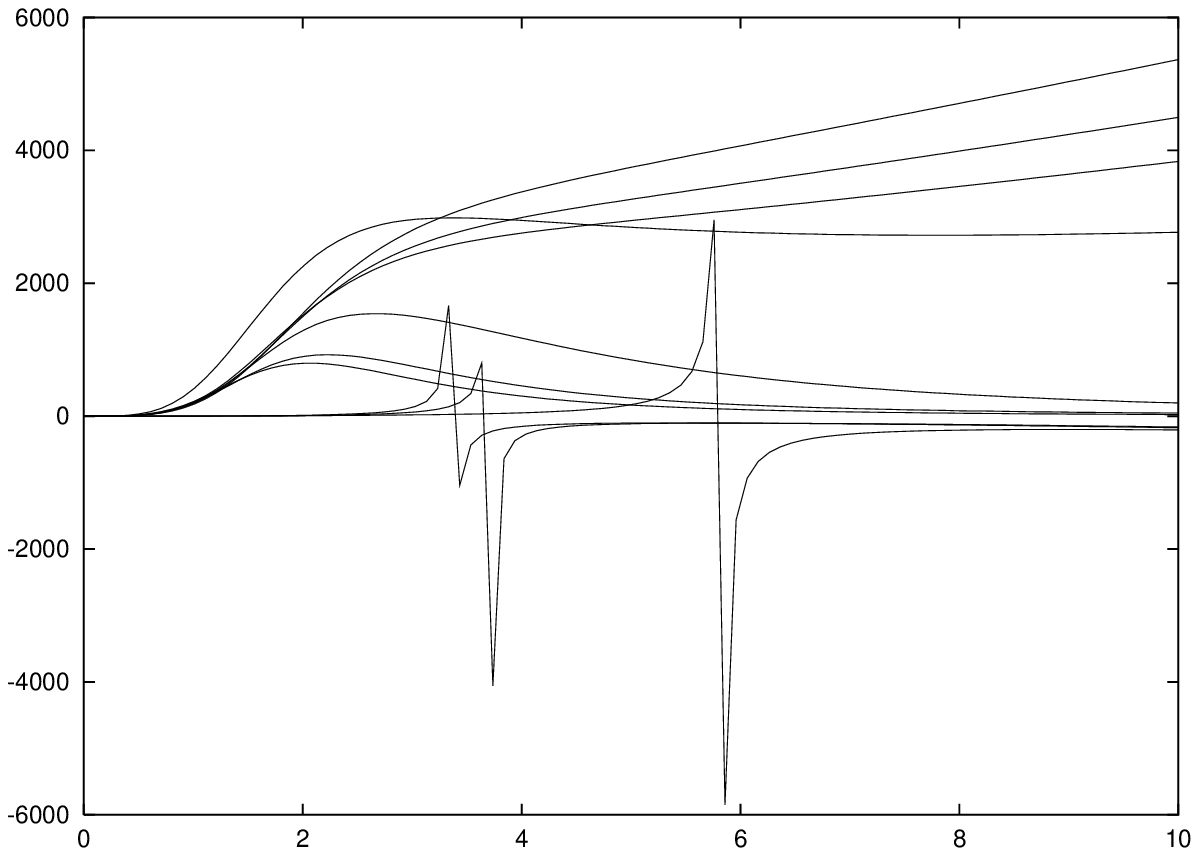} \\
 \tilde r  &  \tilde r \\
(a)  & (b)
\end{array}
$$	
\caption{(a) ${\rm U}_+$ (lower curves),  ${\rm U}_-$ (upper curves) for  
Bonnor-Sackfield-like disks  with $\gamma=1$, $\alpha=1.5$ and $p=q=0.5$,
$1.0$, and $1.5$  and  $\rm{U}^2$ for $\gamma=1$, $\alpha=1.5$ and $b=0.5$,
$1.0$, and $1.5$, as functions of $\tilde r$. $(b)$ $\tilde h^2_+$ (scaled by a
factor of 500),   $\tilde h^2_-$ (sharp curves) for the same values of the
parameters and $\tilde h^2$ (upper curves scaled by a factor of 400) for 
$\gamma=1$, $\alpha=1.5$ and  $b=0.5$, $1.0$, $1.5$ and $4.0$ (bottom curve).} 
\label{fig:bvm}
\end{figure*}

\begin{figure*}
$$
\begin{array}{cc}
\tilde \epsilon _-, \ \tilde \epsilon _+,  \ \tilde \epsilon _ \pm  & \tilde
\sigma _- , \  - \tilde  \sigma _+,   \ -  \tilde \sigma _{e \pm}, \  \mp 
\tilde \sigma _{m \pm}   \\
\epsfig{width=2.5in,file=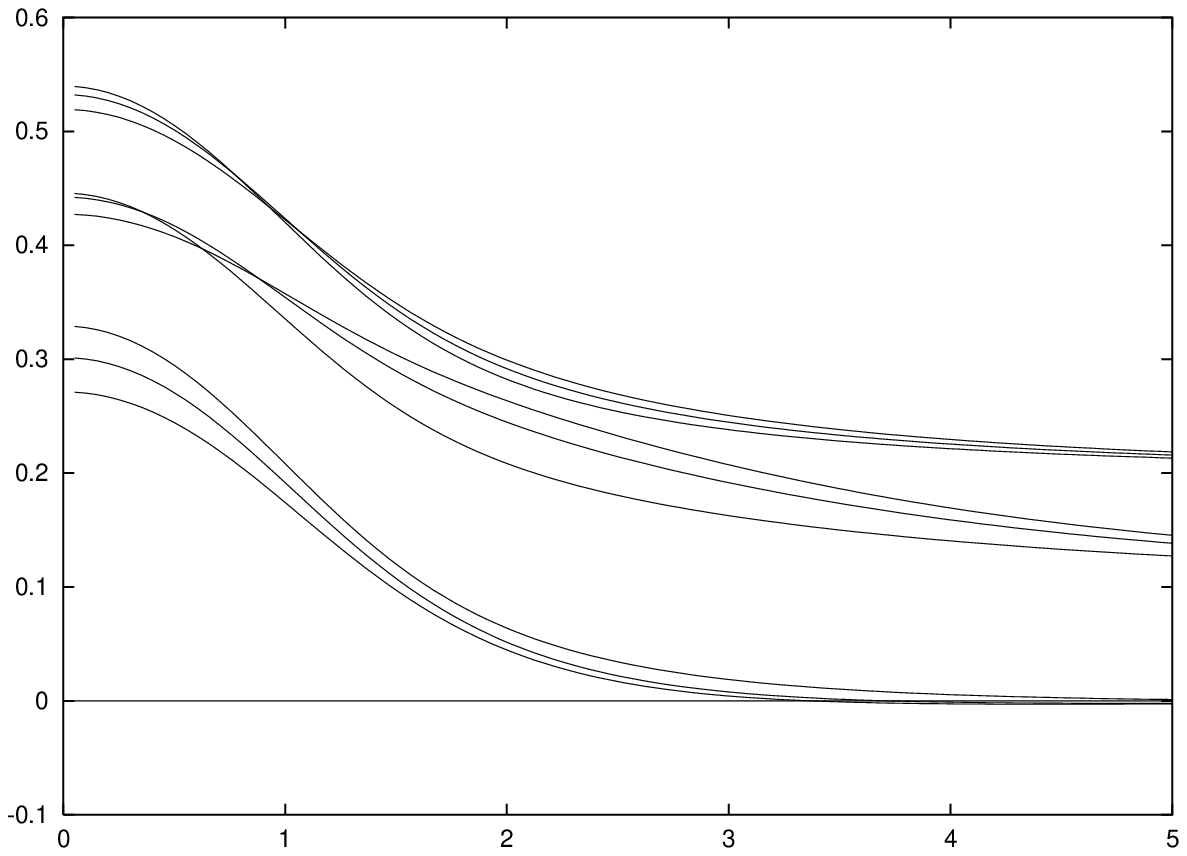} & \epsfig{width=2.5in,file=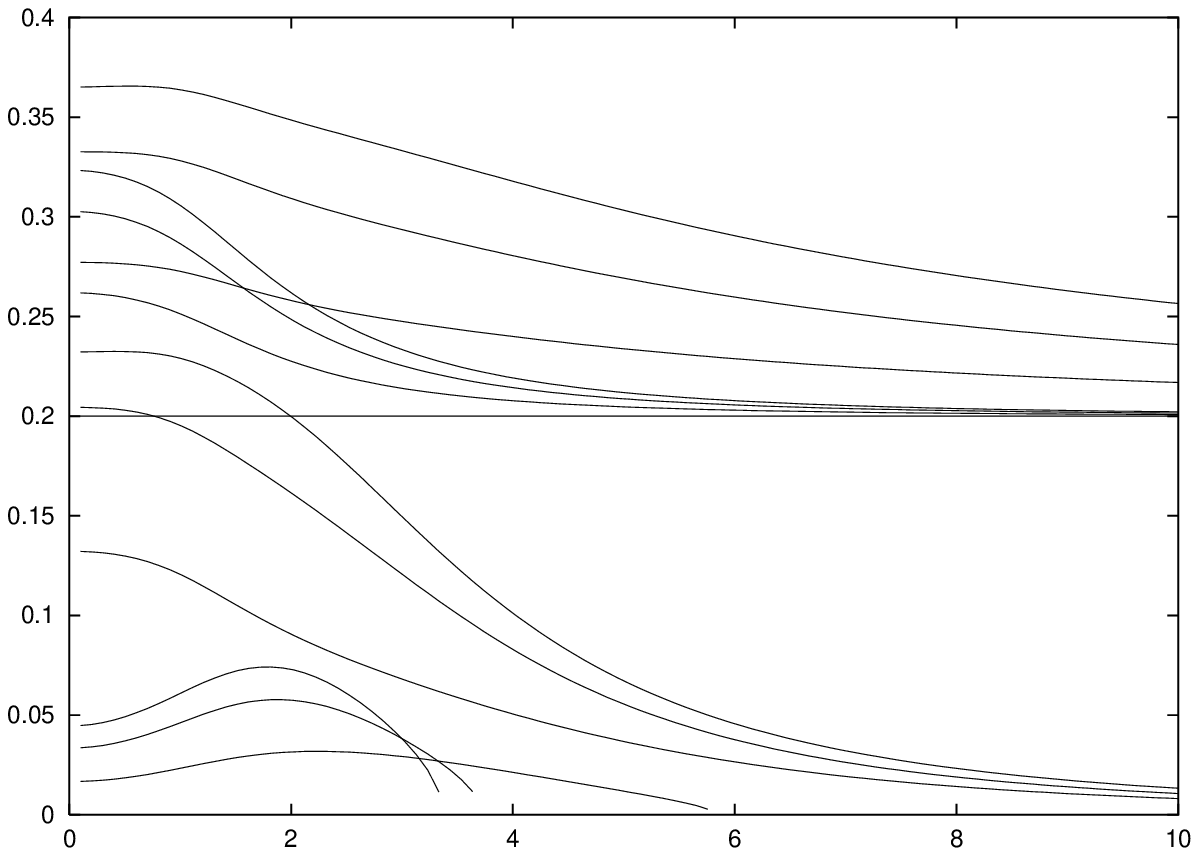}  \\
\tilde r  & \tilde r \\
(a)  & (b)
\end{array}
$$	
\caption{$(a)$ $\tilde \epsilon _-$ (lower curves), $\tilde \epsilon _+$ (moved
upwards a factor of 0.1)  for  Bonnor-Sackfield-like  disks  with $\gamma=1$,
$\alpha=1.5$ and $p=q=0.5$ , $1.0$, and $1.5$ and $\tilde \epsilon _\pm$ 
(upper curves moved upwards a factor of 0.2) for $\gamma=1$, $\alpha=1.5$ and 
$b=0.5$, $1.0$ and  $1.5$.  $(b)$  $\tilde \sigma _-$ (lower curves),   $\tilde
\sigma _+$,    $\tilde \sigma _{e \pm}$  (moved upwards a factor of 0.2) and 
$\tilde \sigma _{m \pm}$ (upper curves moved also upwards a factor of 0.2) for
the same values of the parameters.} \label{fig:becrc}
\end{figure*}

\begin{figure*}
$$
\begin{array}{c}
 - \tilde \sigma _+, \  \tilde \sigma _-     \\
\epsfig{width=2.5in,file=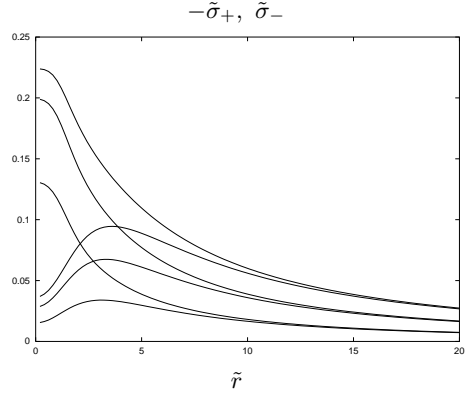}  \\
\tilde r  
\end{array}
$$	
\caption{$\tilde \sigma _+$ (upper curves)  and   $\tilde \sigma _-$  for not
electro-geodesic  Bonnor-Sackfield-like    disks  with $\gamma=1$, $\alpha=1.5$
and $p=q=0.5$ (lower curves), $1.0$, and $1.5$ (upper curves).}
\label{fig:bcng}
\end{figure*}

In Figs. \ref{fig:bepco}-\ref{fig:bcng} the plots of the physical  quantities 
related with the disks  are presented for disks with $\gamma =1$, $\alpha=1.5$
and $p=q=0$, $0.5$, $1.0$, and $1.5$, also as functions of $\tilde{r}$. We see
that these functions have the same  behavior that the previous cases.
Therefore, when we consider  counterrotating  streams moving along
electro-geodesics only the central region of the disks is well behaved, 
whereas in the case of  counterrotating streams not circulating  along
electro-geodesics or for electrostatic (magnetostatic) fields  we obtain disks
with a physically acceptable CRM for many values of the parameters. 

\subsection{CRM for Kerr-like disks}

Finally, other  family of solutions to the Einstein-Maxwell equations, is a
Kerr-like solution which is given by    
\begin{subequations}\begin{eqnarray}
\nu  &=& \ln \left[ \frac{a^2 x^2 - b^2 y^2 - 1}{(a x + 1)^2 - b^2 y^2}
\right], \\
& & \nonumber   \\
\lambda &=& 2 \ln \left[\frac{a^2 x^2 - b^2 y^2 - 1}{a^2 (x^2 - y^2)} \right],
\\
& & \nonumber   \\
A_t &=& \frac {2 \sqrt 2 py}{(ax + 1)^2 - b^2 y^2},  \\
& & \nonumber \\
A_\varphi  &=& -\frac{\sqrt 2 kq (1 - y^2)(ax + 1)}{a(a^2 x^2 - b^2 y^2 - 1)}, 
\end{eqnarray}\label{eq:kerr}\end{subequations}
where $a^2=1+b^2$, with $b^2 = p^2 + q^2 $. Here $p$ and $q$ are also the 
electric and magnetic parameters, respectively. $x$ and $y$ are again the
prolate spheroidal coordinates, given by  (\ref{eq:coorp}). For $b=0$ this
solution  also  reduces to the Darmois metric. Moreover, for  $q=0$ we have a
Kerr-like electrostatic solution. This  solution can   be generated, in these
coordinates,  using a well-know theorem  proposed by Bonnor  (see Ref.
\cite{B2}) from  the Kerr  vacuum solution. And for $p=0$  one  obtains its 
magnetostatic equivalent.  This  solution is asymptotically flat and  was
firts  studied by Bonnor \cite{B3}  in a different  system of coordinates  and 
describes the field of a massive magnetic dipole.

The physical quantities associated with the disks  can now be written as 
\begin{subequations}\begin{eqnarray}
\tilde {\epsilon} & = & \frac{8 a^4 \bar{y}}{(a^2 \bar{x}^2 - b^2 \bar{y}^2 -
1)^2 [(a \bar{x} + 1)^2 - b^2 \bar{y}^2]^2}  \nonumber \\
&&	\nonumber	\\
&& \times \left\{ (\bar{x}^2 - \bar{y}^2)[a(\bar{x}^2 - 1)[(a\bar{x} +
1)^2+b^2\bar{y}^2] -2b^2\bar{x}(a\bar{x}+1)(1-\bar{y}^2)] \right . \\
&&	\nonumber	\\
&& \left . \quad - \ 2 \bar{x} (\bar{x}^2 - 1)(1 - \bar{y}^2) [(a \bar{x} +
1)^2 - b^2 \bar{y}^2] \right\} , \nonumber \\
&&	\nonumber	\\
\tilde {p}_\varphi &=& \frac {16a^4 \bar{x} \bar{y}(\bar{x}^2 - 1 )(1 -
\bar{y}^2)}{(a^2 \bar {x}^2 - b^2\bar{y}^2 - 1)^2 [(a\bar{x} + 1)^2 -
b^2\bar{y}^2]}, \\
&&	\nonumber	\\
\tilde{\rm j}_t &=& - \frac {4 \sqrt 2 a^4 p (\bar{x}^2 - \bar{y}^2)\{\bar{x} (1
- \bar{y}^2)[(a \bar{x} + 1)^2 + b^2 \bar{y}^2] - 2 a \bar{y}^2 (a \bar{x} +
1)(\bar{x}^2 - 1)\}  }{(a^2 \bar{x}^2 - b^2 \bar{y}^2 - 1)[(a \bar{x} + 1)^2 -
b^2 \bar{y}^2]^3},  \\
&& \nonumber	\\
{\rm j}_\varphi &=& - \frac{2 \sqrt 2 a^4 q \bar{y}(\bar{x}^2 - 1)(1 -
\bar{y}^2)(\bar{x}^2 - \bar{y}^2)[(a \bar{x} + 1)(3a\bar{x}+ 1) + b^2
\bar{y}^2]}{(a^2 \bar{x}^2 - b^2 \bar{y}^2 - 1)^3[(a \bar{x} + 1)^2 - b^2
\bar{y}^2]},
\end{eqnarray}\end{subequations}
where $\bar{x}$ and  $\bar{y}$ are given by Eqs. (\ref{eq:xbar}) and
(\ref{eq:ybar}).

\begin{figure*}
$$
\begin{array}{cc}
\tilde \epsilon, \  \tilde p_\varphi  & \tilde {\rm j}_t, \ {\rm j}_\varphi \\
\epsfig{width=2.5in,file=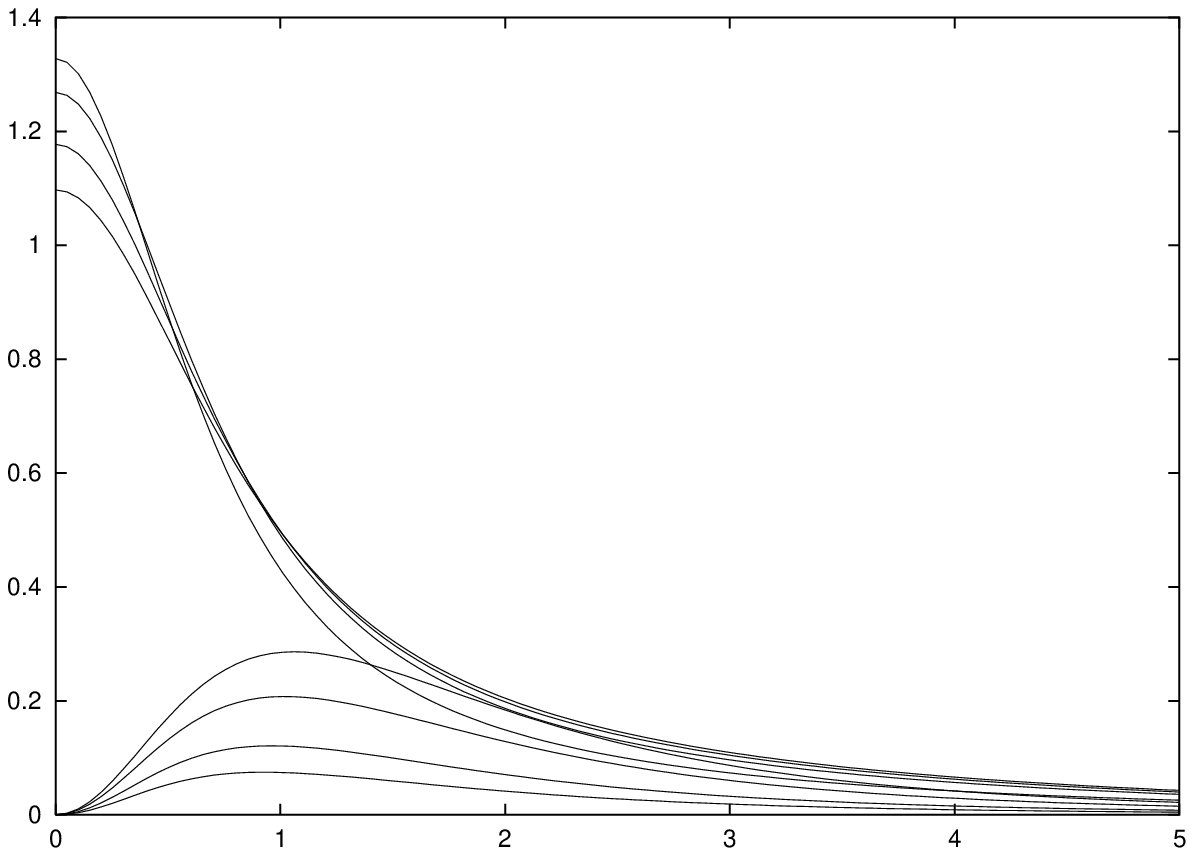} & \epsfig{width=2.5in,file=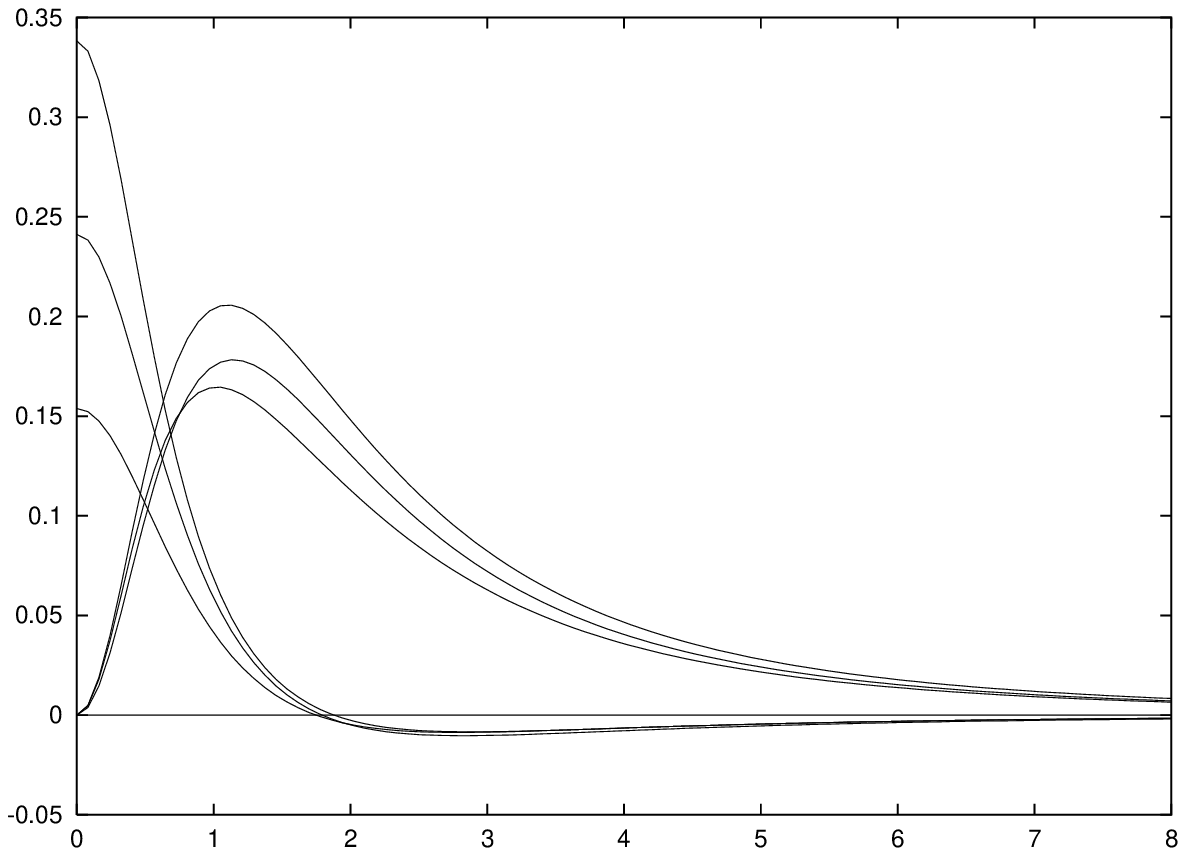} \\
\tilde r & \tilde r    \\
(a)     &   (b)
\end{array}
$$	
\caption{$(a)$ $\tilde \epsilon$ (lower curves ) and $\tilde p_\varphi$,  as
functions of $\tilde r$ for  Kerr-like  disks  with $\alpha=1.7$ and $p=q=0$
(bottom and top curves), $0.4$, $0.7$, and $1.5$ (top and bottom curves). 
$(b)$ $\tilde {\rm j}_t$ (upper curves ) and ${\rm j}_\varphi$ for the same
values  of the parameters.}\label{fig:kepco}
\end{figure*}

\begin{figure*}
$$
\begin{array}{cc}
{\rm U}_+, \ \ -{\rm U}_-, \ \   \rm U^2 & \tilde h^2_+, \ \  \tilde h^2_-, \ \
\tilde h^2   \\
\epsfig{width=2.5in,file=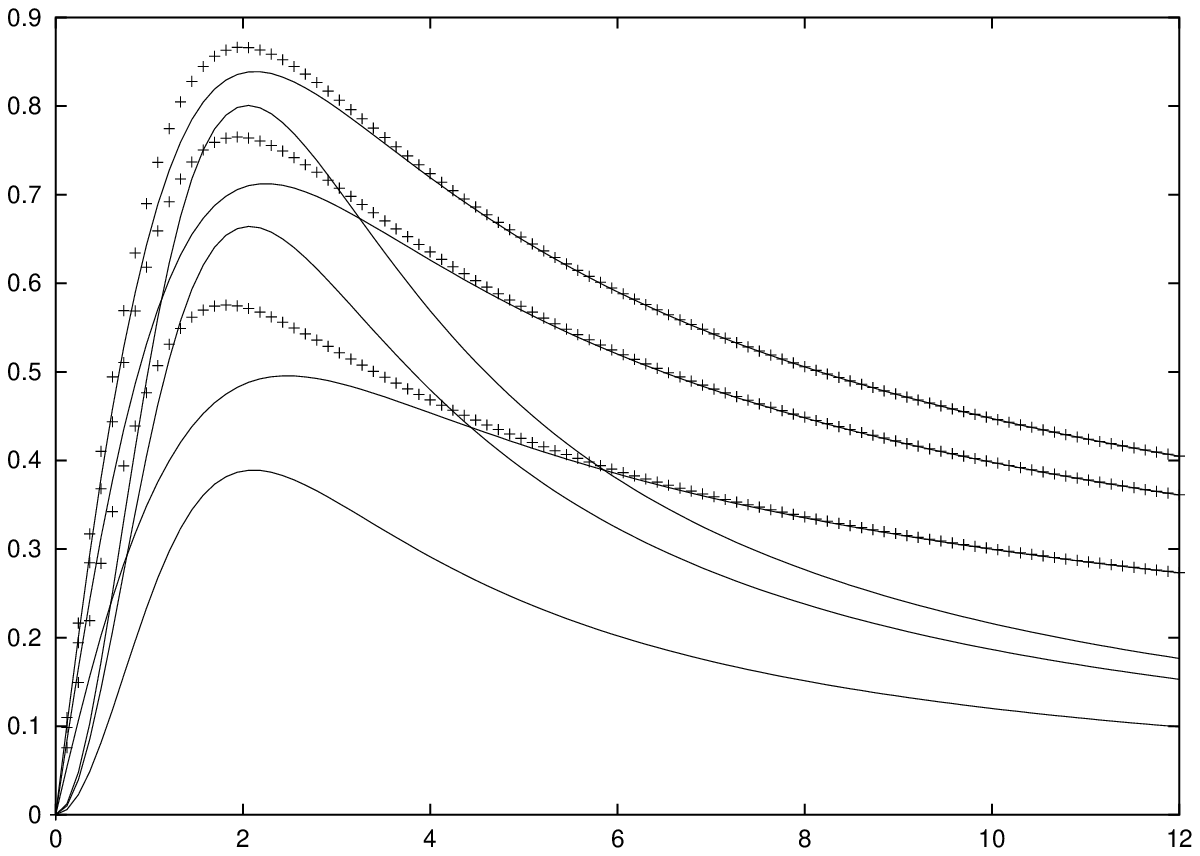} &  \epsfig{width=2.5in,file=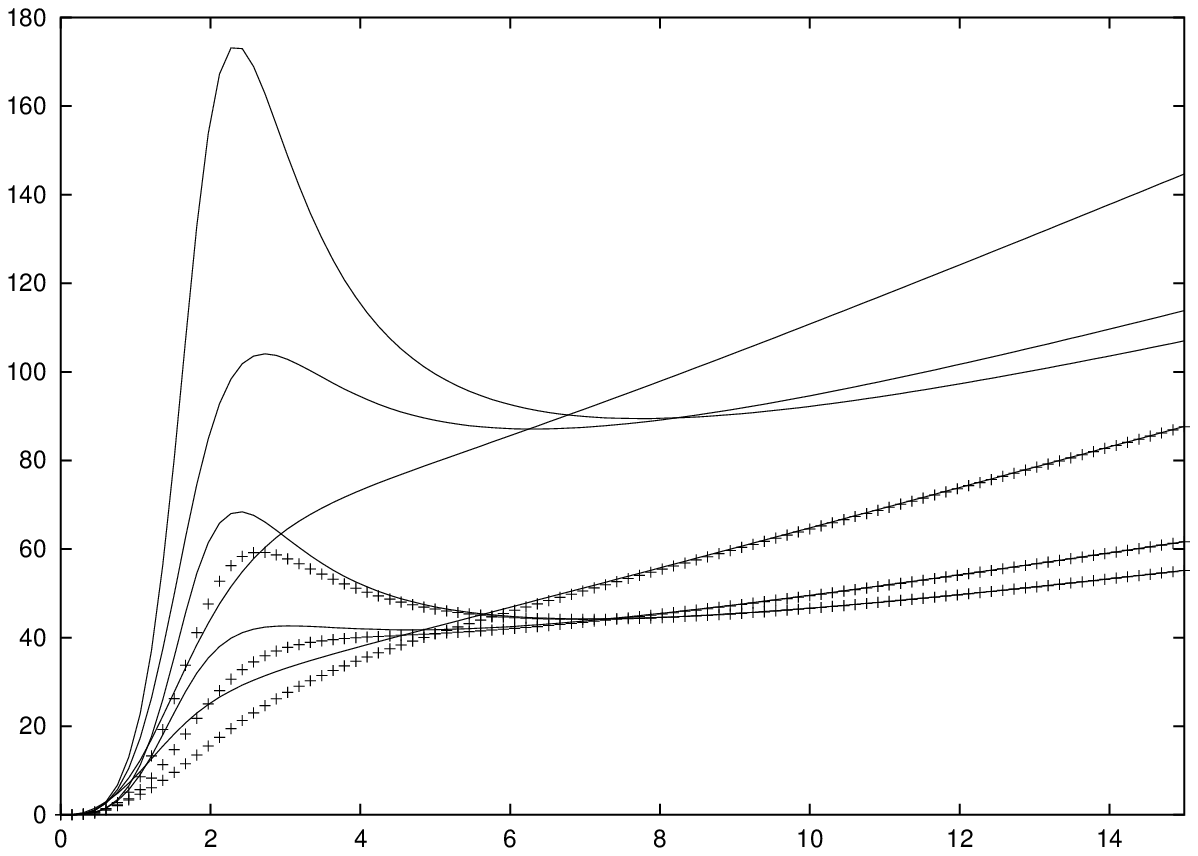} \\
 \tilde r  &  \tilde r \\
(a)  & (b)
\end{array}
$$	
\caption{ (a) ${\rm U}_+$ and  ${\rm U}_-$ (curves with crosses)  for Kerr-like
disks  with  $\alpha=1.7$ and $p=q=0.4$, $0.7$, and $1.5$, and ${\rm U}^2$ for 
$\alpha=1.7$ and $b=0.4$,  $0.7$, $1.5$, as functions of $\tilde r$.  $(b)$
$\tilde h^2_+$ (curves with crosses), $\tilde h^2_-$ for the same values of the
parameters and $\tilde h^2$ (upper curves scaled by a factor of 2) for 
$\alpha=1.7$ and $b=0.4$,  $0.7$, $1.5$ and $4.0$ (bottom curve).}
\label{fig:kvm}
\end{figure*}

\begin{figure*}
$$
\begin{array}{cc}
\tilde \epsilon _-, \ \tilde \epsilon _+,  \ \tilde \epsilon _ \pm  & \tilde
\sigma _- , \  - \tilde  \sigma _+,   \ -  \tilde \sigma _{e \pm}, \  \mp 
\tilde \sigma _{m \pm}   \\
\epsfig{width=2.5in,file=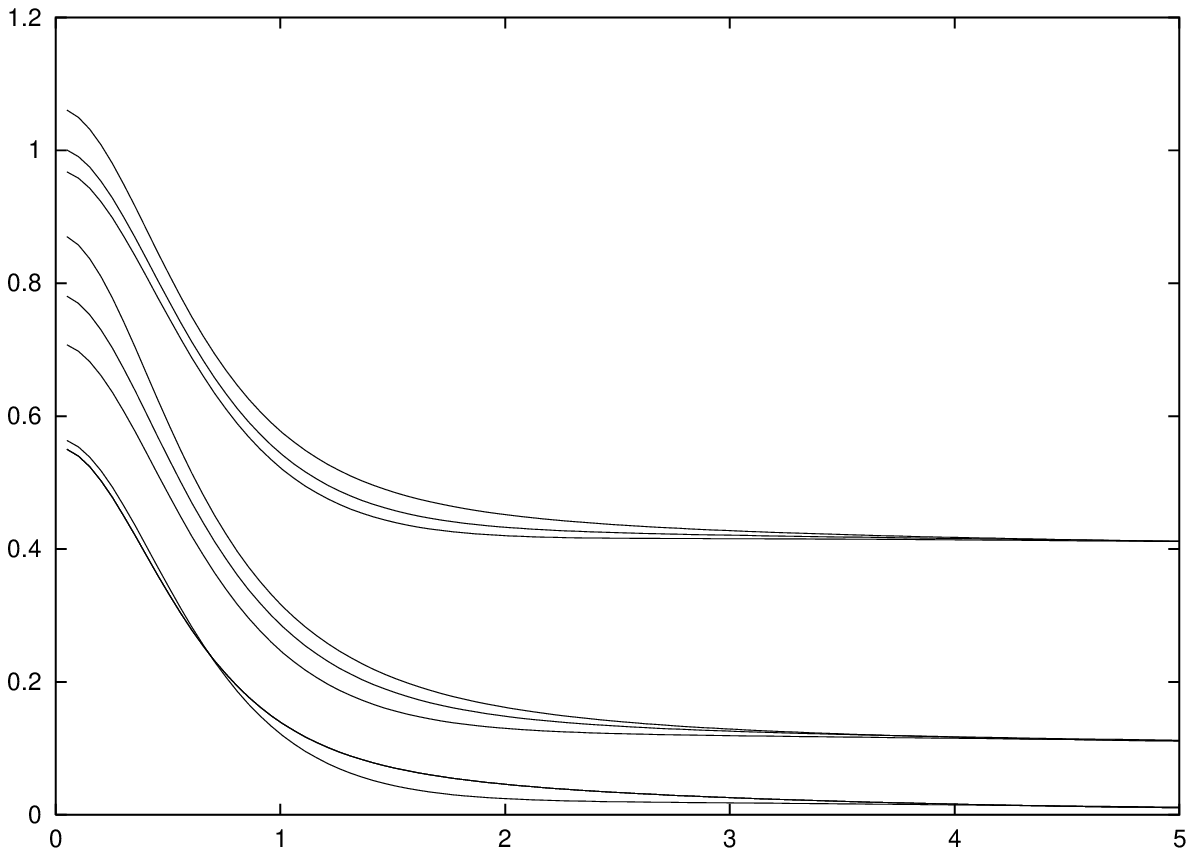} & \epsfig{width=2.5in,file=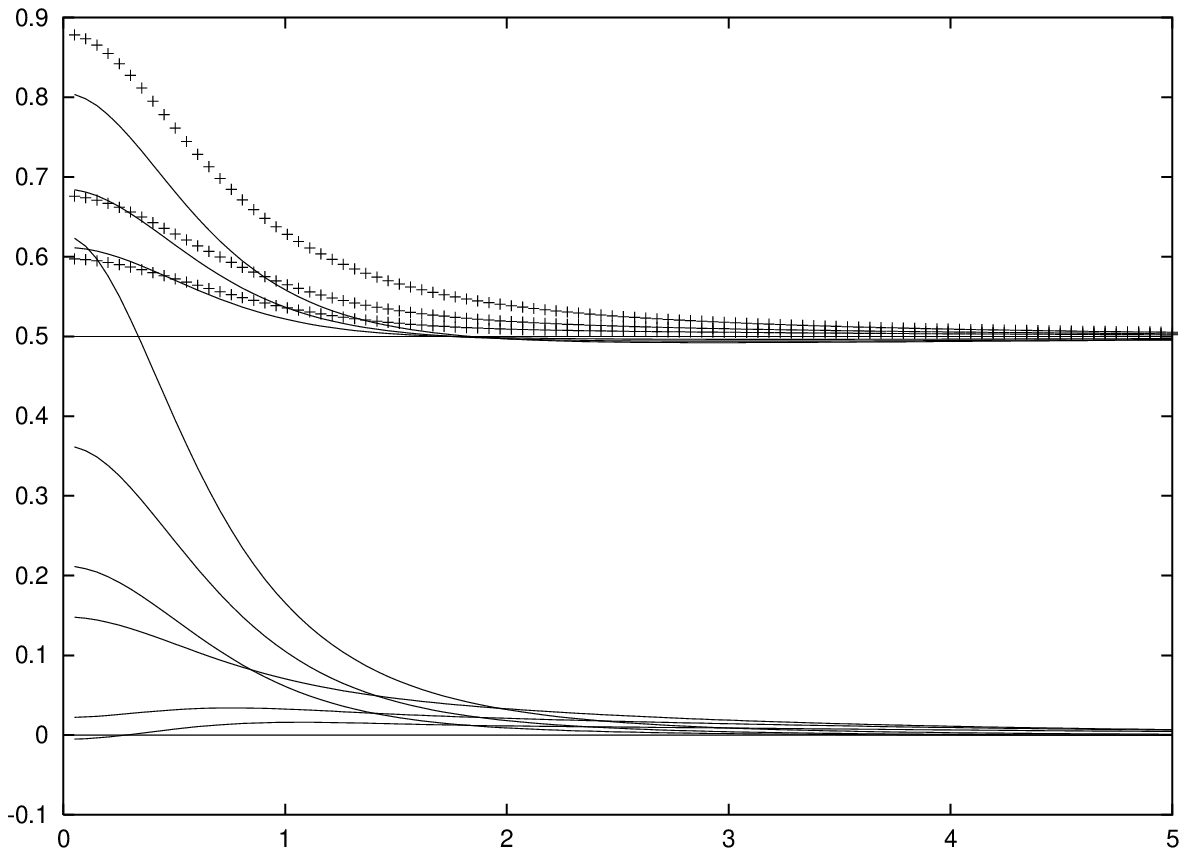}  \\
\tilde r  & \tilde r \\
(a)  & (b)
\end{array}
$$	
\caption{$(a)$ $\tilde \epsilon _-$ (lower curves), $\tilde \epsilon _+$ (moved
upwards a factor of 0.1)  for  Kerr-like  disks  with $\alpha=1.7$ and
$p=q=0.4$, $0.7$, and $1.5$ and $\tilde \epsilon _\pm$  (upper curves moved
upwards a factor of 0.4) for $\alpha=1.7$ and  $b=0.4$, $0.7$ and  $1.5$. 
$(b)$  $\tilde \sigma _-$ (lower curves), $\tilde \sigma _+$, $\tilde
\sigma _{e \pm}$  (moved upwards a factor of 0.5) and  $\tilde \sigma _{m \pm}$
(curves with crosses moved also upwards a factor of 0.5) for the same values of
the parameters.} \label{fig:kecrc}
\end{figure*}

\begin{figure*}
$$
\begin{array}{c}
 - \tilde \sigma _+, \  \tilde \sigma _-     \\
 \epsfig{width=2.5in,file=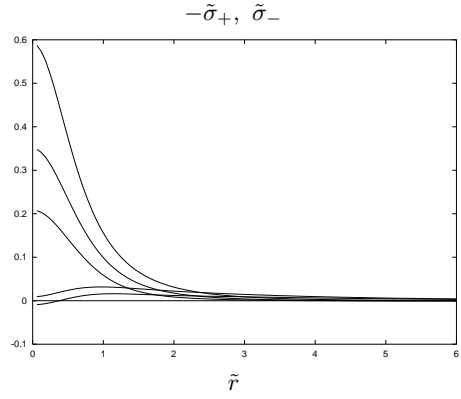}  \\
  \tilde r  
\end{array}
$$	
\caption{$\tilde \sigma _+$ (upper curves)  and   $\tilde \sigma _-$  for not
electro-geodesic  Kerr-like    disks  with $\alpha=1.7$ and $p=q=0.4$ (lower
curves), $0.7$, and $1.5$ (upper curves).} \label{fig:kcng}
\end{figure*}

In figs. \ref{fig:kepco}-\ref{fig:kcng} the plots of the physical quantities
describing the disks  are shown for $\alpha=1.7$ and $p=q=0$, $0.4$, $0.7$, and
$1.5$, as functions of $\tilde{r}$. The energy density  behaves in the opposite
way to the previous cases. That is, near to the disk center it increases when  
the electromagnetic field is applied and decreases later.  The other quantities
have a  similar behavior to the earlier cases. However, the electric charge
density ${\rm j}_0$ after some $\tilde r$  takes  negative values.  Likewise,
the quantities corresponding to the CRM are shown in following figures for the
same values of the  parameters, also as functions of $\tilde{r}$. Again, we
first consider the two counterrotating streams circulating along
electro-geodesics. 

Now the tangential velocities ${\rm U}_+$  and ${\rm U}_-$  are always  less
than the light velocity, falling to zero at infinity (Fig. \ref{fig:kvm}$(a)$).
Note moreover that  the two counterrotating streams initially circulate with
different velocities and  then become equal and opossite. Thus these disks
models  are   well behaved  everywhere on the disks.  We also found that  for
some values of the parameters   cannot be constructed disks from the CRM
because $\rm{U}^2>1$. We also note that  the presence of electromagnetic field 
can stabilize the CRM against radial perturbations (Fig. \ref{fig:kvm}$(b)$). 
On the other hand, the mass densities $\tilde \epsilon _+$, $\tilde \epsilon
_-$ and  $\tilde \epsilon _\pm$  (Fig. \ref{fig:kecrc}$(a)$) are also always
positive and fall to zero at infinity.

In Fig. \ref{fig:kecrc}$(b)$  we have plotted the charge densities $\tilde
\sigma _+$, $\tilde \sigma _-$, $\sigma _{e \pm}$ and $\sigma _{m \pm}$. We see
that  $\tilde \sigma _+$ and  $\sigma _{m \pm}$ are  always positive, falling
to zero at infinity, whereas $\tilde \sigma _-$ can  take initially negative
values and $\sigma _{e \pm}$ become negative after  some value of $\tilde{r}$
and then fall to zero at infinity. Thus, the Kerr-like disk with $\alpha=1.7$
and $p=q=1.5$ can be constructed from the counterrotating streams circulating
along  electro-geodesics.  Finally,  the electric charge densities when the two
streams do not move on electro-geodesics  are shown   in Fig. \ref{fig:kcng},
being always real quantities. Therefore, in all the cases considered  we can 
build Kerr-like disks with a phisically acceptable CRM for many values of the
parameters.   

\section{Discussion}

A detailed study of the counterrotating model for generic electrovacuum static 
axially symmetric relativistic thin disks without radial pressure was
presented. A general constraint over the counterrotating tangential velocities
was found, needed to cast the surface energy-momentum tensor of the disk in
such a way that it can be interpreted as the superposition of two
counterrotating  charged dust fluids. The constraint found is completely
equivalent to the necessary and sufficient condition obtained in reference
\cite{FMP}.  We next showed that this constraint is satisfied if we take the
two counterrotating  fluids as circulating along electro-geodesics. We note
that en general the two fluids circulate with different velocities. However,
when the spacetime is electrostatic (or magnetostatic) the two geodesic fluids
circulate with equal and opposite velocities. We also have obtained explicit
expressions for the energy densities,  current densities and velocities of the
counterrotating streams in terms of the energy density, azimuthal pressure and
planar current density of the disks, that are also equivalent to the
correspondig expressions in reference \cite{FMP}.

Four families of models of counterrotating charged disks were considered in the
present work based in simple solutions to the vacumm Einsteins-Maxwell
equations  in the static axisymmetric case generated by conventional
solution-generating tecniques \cite{KSHM}.  When the two counterrotating
streams move on electro-geodesisc,    Chazy-Curzon-like, Zipoy-Voorhees and
Bonnor-Sackfield-like counterrotating  disks  can be constructed only  with  a
well behaved  central region,  whereas for  Kerr-like fields   we found  for
many values of the parameters  disks  with a physically acceptable CRM. On the
other hand, when we consider the two counterrotating fluids not circulating on
electro-geodesic but with equal and opposite velocities or for electrostatic
(or magnetostatic) fields, we found for many values of the parameter   disks
with a well behaved  CRM everywhere. In the case of  Chazy-Curzon-like,
Zipoy-Voorhees and Bonnor-Sackfield-like fields,  we also   saw that  the  
presence of electric (magnetic) field  can make unstable against radial
perturbations the CRM, and conversely, stabilizes the CRM in the case of
Kerr-like disks.
 
Finally, the generalization of the counterrotating model presented here to the
case  with radial pressure is in consideration. Also, the generalization to 
rotating thin disks with or without radial pressure  in presence of
electromagnetic fields is being considered. 
   
\begin{acknowledgments}

Gonzalo Garc\'\i a R. wants to thank a Fellowship from Vicerrector\'\i a
Acad\'emica, Universidad Industrial de Santander.

\end{acknowledgments}




\end{document}